\documentclass[preprint2,twoside]{myaastex}

 \usepackage{amssymb}                       
 \usepackage{amsmath}                       
\usepackage[ps2pdf,bookmarks=true,breaklinks=true,hypertexnames=false,colorlinks=true,pdfstartview=FitV,linkcolor=red,citecolor=blue,urlcolor=green]{hyperref}
\newcommand{\myemail}{gdangelo@arc.nasa.gov}

\newcommand{\bmath}[1]{\boldsymbol{#1}}

\newcommand{\gvec}[1]{\bmath{#1}}                    
\newcommand{\gOmega}{\bmath{\Omega}}                 
\newcommand{\gdotOmega}{\dot{\bmath{\Omega}}}        
\newcommand{\vdp}{\bmath{r}-\bmath{r}_{\mathrm{p}}}  
\newcommand{\mdp}{|\bmath{r}-\bmath{r}_{\mathrm{p}}|}
\newcommand{\rp}{\mbox{$r_{\mathrm{p}}$}}            
\newcommand{\phip}{\mbox{$\phi_{\mathrm{p}}$}}       
\newcommand{\racc}{\mbox{$r_{\mathrm{acc}}$}}        
\newcommand{\tacc}{\mbox{$\tau_{\mathrm{acc}}$}}     
\newcommand{\refEqt}[1]{Equation~(\ref{#1})}         
\newcommand{\refeqt}[1]{equation~(\ref{#1})}         
\newcommand{\refeqs}[2]{equations~(\ref{#1}) and (\ref{#2})}
\newcommand{\refeqp}[1]{eq.~[\ref{#1}]}              
\newcommand{\refFgt}[1]{Figure~\ref{#1}}             
\newcommand{\refFgp}[1]{Fig.~\ref{#1}}               
\newcommand{\refTab}[1]{Table~\ref{#1}}              
\newcommand{\refSect}[1]{Section~\ref{#1}}           

\newcommand{\refSecp}[1]{\S~\ref{#1}}                
\newcommand{\MSun}{\mbox{$M_{\sun}$}}                
\newcommand{\MStar}{\mbox{$M_{*}$}}                  
\newcommand{\MJup}{\mbox{$M_{\mathrm{J}}$}}          
\newcommand{\Mp}{\mbox{$M_{\mathrm{p}}$}}            
\newcommand{\Md}{\mbox{$M_{\mathrm{D}}$}}            
\newcommand{\dMp}{\dot{M}_{\mathrm{p}}}              
\newcommand{\dMs}{\dot{M}_{*}}                       
\newcommand{\taum}{\tau_{\mathrm{M}}}                
\newcommand{\taue}{\tau_{\mathrm{E}}}                
\newcommand{\Rhill}{\mbox{$R_\mathrm{H}$}}           
\newcommand{\AU}{\mbox{\textrm{AU}}}                 
\newcommand{\Msyr}{\mbox{$\MSun\,\mathrm{yr}^{-1}$}}

\newcommand{\sdunits}{\mbox{$\mathrm{g}\,\mathrm{cm}^{-2}$}}
\received{year Month day}
\revised{year Month day}
\accepted{year Month day}
\ccc{code}
\cpright{none}{73 AC}

\shorttitle{Evolution of Giant Planets in Eccentric Disks}
\shortauthors{D'Angelo, Lubow, \& Bate}

\begin{document}


\title{\textbf{%
       Evolution of Giant Planets in Eccentric Disks\altaffilmark{\dag}%
      }}


\author{\textsc{Gennaro D'Angelo\altaffilmark{1}}}
\affil{School of Physics,
       University of Exeter,
       Stocker Road,
       Exeter EX4 4QL,
       United Kingdom}
\affil{NASA-ARC,
       Space Science and Astrobiology Division,
       MS 245-3, Moffett Field, CA 94035, USA}
\email{\myemail\\[2mm]}
\and
\author{\vspace*{-5mm}\textsc{Stephen H. Lubow}}
\affil{Space Telescope Science Institute,
       3700 San Martin Drive, 
       Baltimore, 
       MD 21218, USA}
\email{lubow@stsci.edu\\[2mm]}
\and
\author{\vspace*{-5mm}\textsc{Matthew R. Bate}}
\affil{School of Physics,
       University of Exeter,
       Stocker Road,
       Exeter EX4 4QL,
       United Kingdom}
\email{mbate@astro.ex.ac.uk\\[-10mm]}



\small

\begin{abstract}
We investigate the interaction between a giant planet and a viscous
circumstellar disk by means of high-resolution, two-dimensional
hydrodynamical simulations.
We consider planet masses that range from $1$ to $3$ Jupiter masses 
($\MJup$) and initial orbital eccentricities that range from $0$ to 
$0.4$.
We find that a planet can cause eccentricity growth in a disk region 
adjacent to the planet's orbit, even if the planet's orbit is circular. 
Disk-planet interactions lead to growth in a planet's orbital eccentricity. 
The orbital eccentricities of a $2\,\MJup$ and a $3\,\MJup$ planet increase
from $0$ to $0.11$ within about $3000$ orbits.
Over a similar time period, the orbital eccentricity of a $1\,\MJup$ 
planet grows from $0$ to $0.02$. For a case of a $1\,\MJup$ planet with 
an initial eccentricity of $0.01$, the orbital eccentricity grows to 
$0.09$ over $4000$ orbits.
Radial migration is directed inwards, but slows considerably as a 
planet's orbit becomes eccentric. If  a planet's orbital eccentricity 
becomes sufficiently large, $e \ga 0.2$, migration can reverse and so 
be directed outwards.
The accretion rate towards a planet depends on both the disk and the planet 
orbital eccentricity and is pulsed over the orbital period. Planet mass 
growth rates increase with planet orbital eccentricity. For $e \sim 0.2$ 
the mass growth rate of a planet increases by $\sim 30\%$ above the value 
for $e=0$.  For $e \ga 0.1$, most of the accretion within the planet's 
Roche lobe  occurs when the planet is near the apocenter. Similar accretion 
modulation occurs for flow at the inner disk boundary which represents 
accretion toward the star.
\end{abstract}


\keywords{accretion, accretion disks ---
          hydrodynamics ---
          methods: numerical ---
          planetary systems: formation}

\section{Introduction}\symbolfootnote[0]{\hspace*{-0.5em}$^\dag$%
                      To appear in %
                      \textsc{The Astrophysical Journal} %
                      (v652 n2 December 1, 2006 issue).
                      Also available as ApJ preprint doi: %
                      \texttt{10.1086/508451}.
                                        }
\label{sec:introduction}
\footnotetext[1]{NASA Postdoctoral Fellow.}
\setcounter{footnote}{1}
\thispagestyle{empty}
A striking property of the extrasolar planets known to date is the
range of their orbital eccentricities, which is far wider than that
of planets in the solar system.\footnote{See 
the \textit{California \& Carnegie Planet Search} at
\url{http://exoplanets.org} and
the \textit{Extrasolar Planets Encyclopaedia} at
\url{http://exoplanet.eu}. 
}
Eccentricities are typically $\sim 0.2$--$0.3$, but very low and high 
values are also found \citep[e.g.,][]{marcy2005}. A variety of explanations 
have been proposed to explain the eccentricities. The high eccentricity 
cases may owe their explanation to the presence of a distant binary 
companion \citep{wu2003,takeda2005}. In fact, the large eccentricity 
($e=0.93$) of HD~80606b could be the result of a three-body interaction 
process, known as ``Kozai cycle'' \citep{wu2003}.
Moreover, numerical experiments show that the observed distribution of 
orbital eccentricities for $e\gtrsim 0.6$ can be reproduced by assuming 
the action of a Kozai-type perturbation \citep{takeda2005}.

For more typical eccentricities, other processes are likely to be at 
work. Planet-planet interactions involving scattering on dynamical
timescales is a possibility \citep{rasio1996}. However, numerical 
experiments indicate that interactions between equal mass planets 
would produce more isolated planets on low-eccentricity orbits than 
those observed \citep{ford2001}. Secular interactions between planets 
is another possible means of eccentricity excitation \citep{juric2005}. 
This mechanism assumes that the planets have appropriate initial 
separations for evolution  to occur on secular timescales ($\sim 10^{10}$ 
years). 

Disk-planet interactions can also give rise to planetary eccentricities
\citep{gt1980,pawel1992,goldreich2003}. The evolution of orbital 
eccentricity depends on a delicate balance between Lindblad and corotation 
resonances. If the planet is massive enough to clear a gap, Lindblad 
resonances promote eccentricity growth, while corotation resonances damp 
eccentricity. If the corotation resonances operate at maximal efficiency, 
they dominate over Lindblad resonances by a slight margin, and the result 
is eccentricity damping \citep[][]{gt1980}. Two mechanisms have been 
proposed to weaken the effects of corotation resonances and thereby provide 
eccentricity growth. The first mechanism relies on a large enough gap about 
a planet's orbit to exclude the corotation resonances from the disk, while 
leaving a remaining Lindblad resonance, the 1:3 outer resonance. This 
mechanism requires a massive enough companion and/or low enough viscosity 
to open a wide enough gap. Such a mechanism has been demonstrated for binary 
stars \citep{pawel1991}.
The second mechanism relies on the delicate nature of corotational resonances 
in their ability to weaken (saturate) when the disk viscosity is sufficiently 
small and the planet eccentricity sufficiently large 
\citep[][]{ogilvie2003,goldreich2003}. A small finite initial eccentricity, 
typically $e\sim 0.01$, is required for the eccentricity to grow by this 
mechanism. Moreover, a companion object on a circular orbit can drive 
eccentricity in a circumstellar disk. This process is believed to occur 
in disks involving $10$--$20$ Jupiter-mass ($\MJup$) companions 
\citep{papa2001} and binary stars \citep{lubow1991a,lubow1991b}. The disk 
transfers eccentricity to the planet's orbit by exchanging energy and angular 
momentum. The disk-planet system undergoes a coupled eccentricity evolution.

One goal of this paper is to determine whether the eccentricity growth by 
disk-planet interactions occurs. Simulations by \citet{papa2001} suggested 
that orbital eccentricity is excited when the mass of the embedded body is 
larger than about $10$--$20$ Jupiter-masses, while lower mass companions 
experienced eccentricity damping. By applying higher resolution and 
simulating over longer timescales, we aim to see whether eccentricity growth 
can occur for lower mass, planetary mass, companions. In addition, we are 
interested in the effects of eccentricity on planet migration.

Accretion of gas onto a planet is likely affected by the planet's orbit 
eccentricity. Circumbinary disks surrounding  eccentric orbit binaries
undergo pulsed accretion on  orbital timescales \citep{pawel1996}. A similar 
process could occur with eccentric orbit planets. The mean accretion rate 
could also be modified by the orbital eccentricity. Another goal of this 
paper is to investigate this accretion process.

\pagestyle{myheadings}
\markboth{\hfill Evolution of Giant Planets in Eccentric Disks \hfill}%
       {\hfill \textsc{G. D'Angelo, S. Lubow, \& M. Bate} \hfill}
We apply high-resolution hydrodynamical simulations to investigate 
disk-planet interactions over several thousand orbital periods. In 
\refSect{sec:model_description} we describe the physical model. The 
numerical aspects of the calculations are detailed in 
\refSect{sec:numerical_issues}. Results on the growth of the disk 
eccentricity for a fixed planet orbit are presented in 
\refSect{sec:disk_eccentricity}. Similar results on the growth of 
disk eccentricity were recently reported by \citet{kley2006}. We 
describe results on the planet's orbital evolution in 
\refSect{sec:orbital_evolution}. Results on the pulsed accretion 
are described in \refSect{sec:planet_accretion}. Finally, our 
conclusions are given in \refSect{sec:conclusions}

\section{Model Description}
\label{sec:model_description}
\subsection{Evolution Equations}
\label{sec:evolution_equations}
We assume that the disk matter and planet are in coplanar orbits. In 
order to describe the dynamical interactions between a circumstellar 
disk and a giant planet, we approximate the disk as being two-dimensional,
by ignoring dynamical effects in the direction perpendicular to the orbit 
plane (vertical direction). This approximation is justified by the 
fact that the disk thickness is smaller than the vertical extent of 
the planetary Roche lobes for the cases we consider. Comparisons between 
two- and three-dimensional models of Jupiter-mass planets embedded in 
a disk indicate that the two-dimensional geometry provides a sufficiently 
reliable description of the system 
\citep{kley2001,gennaro2003b,bate2003,gennaro2005}.

We employ a cylindrical coordinate frame $\{O; r, \phi, z\}$ in which the 
disk lies in the plane $z=0$ and the origin, $O$, is located on the primary. 
The reference frame rotates about the disk axis (i.e., the $z$-axis) with 
an angular velocity $\Omega$ and an angular acceleration $\dot{\Omega}$. 
The set of continuity and momentum equations, describing the evolution of 
the disk, is written in a conservative form \citep[see, e.g,][]{gennaro2002} 
and is solved for the surface density, $\Sigma$, and the velocity field
of the fluid, $\gvec{u}$.
The turbulent viscosity forces in the disk are assumed to arise from a 
standard viscous stress tensor for a Newtonian fluid with a constant 
kinematic viscosity, $\nu$, and a zero bulk viscosity 
\citep[see][for details]{m&m}.

A locally isothermal equation of state is adopted by requiring that the
vertically integrated pressure is equal to $p=c^2_{\mathrm{s}}\,\Sigma$ 
and that the sound speed, $c_{\mathrm{s}}$, is equal to the disk aspect 
ratio, $H/r$, times the Keplerian velocity. In this study, we use a 
constant aspect ratio throughout the disk.

The gravitational potential of the disk, $\Phi$, is given by
\begin{equation}
 \label{eq:phi}
 \Phi=%
     - \frac{G\,\MStar}{|\gvec{r}|}%
     - \frac{G\,\Mp}{\sqrt{\mdp^2+\varepsilon^2}}%
     + \frac{G\,\Mp}{|\gvec{r}_{\mathrm{p}}|^3}%
     \,\gvec{r}\bmath{\cdot}\gvec{r}_{\mathrm{p}}\,,
\end{equation}
where $\MStar$ is the mass of the central star whereas $\Mp$ and 
$\gvec{r}_{\mathrm{p}}$ are the mass and the vector position of the planet, 
respectively. The length $\varepsilon$ represents a softening length that 
is needed to avoid singularities in the gravitational potential of the 
planet. The third term on the right-hand side of \refeqt{eq:phi} accounts 
for the acceleration of the origin of the (non-inertial) coordinate frame 
caused by the planet. We ignore disk self-gravity. For the disks we consider, 
the Toomre-$Q$ parameter never drops below $4$ during the simulations.

The orbit of the planet evolves under the gravitational forces exerted by 
the central star and the disk material. Non-inertial forces arising from 
the rotation of the reference frame also need to be taken into account. 
Therefore, the equation of motion of the planet reads
\begin{eqnarray}
 \label{eq:pme}
 \ddot{\gvec{r}}_{\mathrm{p}} &=&
     -\frac{G(\MStar+\Mp)}{|\gvec{r}_{\mathrm{p}}|^3}\,\gvec{r}_{\mathrm{p}}%
     -\gOmega\bmath{\times}%
      \left(\gOmega\bmath{\times}\gvec{r}_{\mathrm{p}}\right)%
     -2\,\gOmega\bmath{\times}\dot{\gvec{r}}_{\mathrm{p}}\nonumber\\%
                              & &
     -\gdotOmega\bmath{\times}\gvec{r}_{\mathrm{p}}
     +\bmath{\mathcal{A}}_{\mathrm{p}}-\bmath{\mathcal{A}}_{*}\,,
\end{eqnarray}
where the angular velocity and acceleration vectors of the rotating frame 
are defined as $\gOmega=\Omega\,\bmath{\hat{z}}$ and 
$\gdotOmega=\dot{\Omega}\,\bmath{\hat{z}}$, respectively. The second, 
third, and fourth terms on the right-hand side of the equation are the 
centrifugal, Coriolis, and angular accelerations, respectively. 

The angular velocity of the reference frame relative to a fixed frame, 
$\Omega=\Omega(t)$, is chosen so as to compensate for the azimuthal 
motion of the planet. Details of how this is achieved can be found in 
\citet{gennaro2005}. Since the azimuthal position of the planet is 
constant, a circular orbit reduces to a fixed point in the rotating frame. 
For a planet orbit with $e>0$, the planet radially oscillates  between 
the pericenter distance, $a(1-e)$, and the apocenter distance, $a(1+e)$. 
In this frame, the trajectory is a straight line of length $2ae$ with 
center at $r=a$. This scheme has the advantage that the planet does not 
drift across the grid in the azimuthal direction. It always maintains
a symmetric azimuthal position with respect to the zone centers (where 
eqs.~[\ref{eq:phi}], [\ref{eq:A2}], and [\ref{eq:A1}] are evaluated) 
while it moves radially. This method helps reduce  artificially unbalanced 
forces  that act on the protoplanet
\citep[see discussion in][]{anelson2003a,anelson2003b}.

The last two terms in \refeqt{eq:pme}, which represent the forces per 
unit mass exerted by the disk material on the planet and the star, are
\begin{equation}
 \label{eq:A2}
 \gvec{\mathcal{A}}_{\mathrm{p}}=%
 G\!\int_{\Md}\!\frac{\left(\vdp\right)\,\mathrm{d}\Md(\gvec{r})}%
                     {\left(\mdp^2 + \varepsilon^2\right)^{3/2}}
\end{equation}
and
\begin{equation}
 \label{eq:A1}
 \gvec{\mathcal{A}}_{*}=%
 G\!\int_{\Md}\!\frac{\gvec{r}\,\mathrm{d}\Md(\gvec{r})}{r^{3}}\,.
\end{equation}
These two vectors are included in \refeqt{eq:pme} only when the disk 
back-reaction is accounted for and the protoplanet is allowed to adjust 
its trajectory in response to the disk torques. When these terms are 
applied, the integration in \refeqs{eq:A2}{eq:A1} is performed over the 
disk mass comprised in the simulated domain, $\Md$.

\subsection{Physical Parameters}
\label{sec:physical_par}
In these calculations, the stellar mass, $\MStar$, is the unit of mass 
and the initial semi-major axis of the planet's orbit, $a_{0}$, represents 
the unit of length. The unit of time, $t_{0}$, is defined such that 
$1/t_{0}=\sqrt{G\,(\MStar+\Mp)/a^{3}_{0}}$.  
When we refer to the ``orbital period'' or ``orbit'' as length of time, 
we actually mean the initial orbital period $t_{0}/(2\pi)$.
To provide estimates of various quantities in physical units, we adopt 
$\MStar=1\,\MSun$ and $a_{0}=5.2\,\AU$, thus one orbit is $\approx11.9$ 
years.

In order to treat the strong perturbations induced by giant protoplanets 
and limit the influence of the imposed radial boundaries, we considered an
extended disk whose radial borders are at $r_{\mathrm{min}}=0.3\,a_{0}$ and 
$r_{\mathrm{max}}=6.5\,a_{0}$. The disk extends over the entire  $2\pi$ radians in 
azimuth around the star. In physical units, the disk models cover a region 
from $1.56$ to $33.8\,\AU$.
The mass of the disk, in the absence of the planet, is 
$\Md=1.58\times10^{-2}\,\MStar$ within the radial limits of the simulated 
region. We used a constant disk aspect ratio, $H/r=0.05$. The unperturbed 
initial surface density is axisymmetric and scales as $r^{-1/2}$, which 
produces an unperturbed density, at $r=a_{0}$, equal to $75.8\,\sdunits$.
However, given the large planetary masses considered in this investigation, 
we also included an initial gap along the orbit of the planet that accounts 
for an approximate balance between viscous and tidal torques 
\citep[e.g.,][]{lin1986}. 
The initial gap profile is based on equation~(5) of \citet{lubow2006}.
The initial gap width is modified by a factor of $1+a_{0} e_{0}$, in order
to account for the planet's initial orbital eccentricity, $e_{0}$.

The initial velocity field in the disk is a Keplerian one that is centered 
on the star and  corrected for the rotation of the frame of reference.
In order to account for the effects due to turbulence in the disk, a 
constant kinematic viscosity, $\nu$, was used. In terms of the Shakura \& 
Sunyaev parameter \citep{S&S1973}, we have 
$\alpha=\alpha_{0}\,\left(a_{0}/r\right)^{1/2}$, where
$\alpha_{0}=4\times10^{-3}$ (in physical units, 
$\nu\simeq10^{15}\,\mathrm{cm}^2\,\mathrm{s}^{-1}$).
Although spatial variations and time fluctuations consistent with the MHD 
turbulence are not included, this relation yields a magnitude of $\alpha$ 
that is in the range found in MHD simulations 
\citep{papa2003,winters2003,rnelson2004}.
The influence of viscosity was explored by performing a few calculations 
with other $\alpha_{0}$ values  ($1.2\times10^{-3}$ and $1.2\times10^{-2}$).

\begin{deluxetable}{ccccccc}
\tabletypesize{\normalsize}
\tablecaption{\small Initial orbital eccentricities.\label{tbl:eccentricities}}
\tablewidth{0pt}
\tablehead{%
\colhead{$\Mp$}  & \multicolumn{6}{c}{$e_{0}$}\\
\cline{2-7}
\colhead{(\MJup)}& \colhead{$0$} & \colhead{$0.01$} & \colhead{$0.1$} & \colhead{$0.2$} & \colhead{$0.3$} & \colhead{$0.4$}%
}
\startdata
   $1$   & $\bullet$ & $\bullet$ & $\bullet$ & $\bullet$ & $\bullet$ & $\bullet$ \\
   $2$   & $\bullet$ &           &           &           &           &          \\
   $3$   & $\bullet$ &           & $\bullet$ & $\bullet$ & $\bullet$ & $\bullet$
\enddata
\end{deluxetable}

Three planetary masses were considered: $1\,\MJup, 2\,\MJup, $ and $3\,\MJup$
(i.e., the mass ratio $q=\Mp/\MStar$ ranges from $1\times10^{-3}$ to 
$3\times10^{-3}$). The planets were set on initially circular or eccentric 
orbits about one solar mass stars. We examined configurations with initial 
eccentricities, $e_{0}$, up to $0.4$. A complete list is given in 
\refTab{tbl:eccentricities}.

At time $t=0$ the planet starts from the pericenter position, while its 
azimuth, $\phip$, remains constant (in the rotating frame) and equal to 
$\pi$ throughout the calculation.
In order to allow the disk to adjust to the presence of the planet, we 
impose two stages to the evolution. During the first phase, the planet's 
orbit is static and terms (\ref{eq:A2}) and (\ref{eq:A1}) are not included 
in \refeqt{eq:pme}. During the second phase, the protoplanet is ``released'' 
from the fixed orbit and is allowed to react to the disk torques via 
the full form of \refeqt{eq:pme}. In the models presented here the first 
phase lasts until the release time, $t=t_{\mathrm{rls}}$, which ranges 
from $1000$ to $1200$ orbits. The second phase lasts from several hundred 
to several thousand orbits.

The value adopted for smoothing radius $\varepsilon$ (in eqs.~[\ref{eq:phi}]
and [\ref{eq:A2}]) resulted from numerical experiments in each orbital 
eccentricity configuration. The chosen value of $\varepsilon$ was the 
smallest that prevented the integration time-step of the hydrodynamics 
equations from getting shorter than $\sim 10^{-6}$ orbits. In models 
involving $1\,\MJup$ and $2\,\MJup$ planets, we set 
$\varepsilon=0.1\,\Rhill$, where $\Rhill=\rp\,\left(q/3\right)^{1/3}$
\citep{bailey1972} is planet's Hill radius (or sometimes called Roche 
radius). In models involving $3\,\MJup$ planets, we applied softening 
lengths between $0.12\,\Rhill$ and $0.2\,\Rhill$. The latter value was 
used at the highest initial orbital eccentricities, $e_{0}=0.3$ and $0.4$. 
We found that torques within the Roche lobe do not dominate the planet 
orbital evolution. Moreover, the smoothing radius does not significantly 
affect planet accretion. Thus, $\varepsilon$ does not likely play an 
important role in these calculations.

For simulations that account for the disk torques on the planet, an 
additional approximation is made, which is described at the beginning
of \refSect{sec:orbital_evolution}.

\section{Numerical Method}
\label{sec:numerical_issues}
The equations of motion of the disk are solved numerically by means 
of a finite-difference scheme that uses a directional operator splitting 
procedure. The method is second-order accurate in space and semi-second-order 
in time \citep{ziegler1997}. Hydrodynamic variables are advected 
by means of a transport scheme that uses a piecewise linear reconstruction 
of the variables with a monotonised slope limiter \citep{vanleer1977}.
High numerical resolution in an extended region around the planet is 
achieved by using a nested-grid technique 
\citep[see][for details]{gennaro2002,gennaro2003b} with fully nested 
subgrid patches, whereby each subgrid level increases the resolution 
by a factor $2$ in each direction.
Tests on the behavior of the nested-grid technique applied in a reference 
frame rotating at a variable rate $\Omega=\Omega(t)$ are reported in the 
Appendix of \citet{gennaro2005}. 
The equation of motion of the planet (\refeqp{eq:pme}) is solved by using
a high-accuracy algorithm described in \citet{gennaro2005}.

\subsection{Grid Resolution}
\label{sec:grid_setup}
\begin{deluxetable}{ccccc}
\tablecaption{\small Grid systems used in the simulations.\label{tbl:grids}}
\tablewidth{0pt}
\tablehead{%
\colhead{Grid} & \colhead{GS1} & \colhead{GS2} & \colhead{GS3}
        & \colhead{GS4}                   \\
\colhead{level}& \colhead{$N_{r}\times N_{\phi}$} & \colhead{$N_{r}\times N_{\phi}$} & \colhead{$N_{r}\times N_{\phi}$} & \colhead{$N_{r}\times N_{\phi}$}
}
\startdata
   $1$  & $313\times 317$          & $313\times 317$          & $623\times 629$         & $623\times 629$       \\
   $2$  & $264\times 264$          & $264\times 264$          & $524\times 524$         &                       \\
   $3$  &                          & $404\times 404$          & 
        &                       \\
\enddata
\tablecomments{\small%
 The grid system GS1 achieves the highest resolution
 $(\Delta r/a_{0}, \Delta \phi)=(0.01,0.01)$ in the region
 $(r,\phi)\in[0.4,3.0]\,a_{0}\times 0.83\pi$, which is azimuthally
 centered on the planet.
 The grid system GS3 resolve the same region with a resolution
 $(\Delta r/a_{0}, \Delta \phi)=(0.005,0.005)$.
 The latter resolution is obtained with the grid system GS2 in the
 region $(r,\phi)\in[0.5,2.5]\,a_{0}\times 0.65\pi$
 (also centered on the planet in azimuth).
}
\end{deluxetable}

The excitation or damping of a protoplanet's orbital eccentricity depends 
on a delicate balance between Lindblad and corotation resonances
\citep[and references therein]{ogilvie2003,goldreich2003}. To study this 
balance, it is necessary to resolve the width of all the resonances 
involved in the process. The locations of first order eccentric Lindblad 
resonances reside in a region that ranges radially from approximately 
$0.6\,a$ to $1.6\,a$. Furthermore, calculations on the saturation of 
isolated  noncoorbital corotation resonances, performed by 
\citet{masset2004}, suggest 
that a minimum resolution requirement to avoid spurious damping of 
eccentricity is that
\begin{equation}
 \label{eq:minres}
 \Delta r/a \lesssim 4.1\,\sqrt{C^{\pm}_{k}\,k\,e\,q}\,,
\end{equation}
where $k$ is the azimuthal wavenumber and $C^{\pm}_{k}$ are coefficients of 
order unity given in \citet{ogilvie2003}. 
Although nested grids provide a linear resolution $\le 0.01$ in a region
$\sim 2.5\,a\times 0.8\pi$, the effects of these corotational 
resonances are not localized in azimuth, so the nested grids do
not substantially improve their overall resolution.
Calculations that follow the orbital evolution of the planet were executed 
with grid systems GS1 and GS2 (see \refTab{tbl:grids} for a description
of all grid systems employed in this study).
According to \refeqt{eq:minres}, the global radial resolution we apply 
$\Delta r/a=0.02$ could produce spurious damping in the outer disk for 
orbital eccentricities $e \la 3\times 10^{-5}/(k\, q)$, with $k >1$.
In the $1\,\MJup$ case, simulations with $0.01 < e < 0.015$ could undergo 
spurious eccentricity damping for the most distant outer resonance $k=2$ 
only. For $e > 0.015$, there is no spurious damping. In the $3\,\MJup$ case, 
simulations with $0.003 < e < 0.005$ could undergo spurious damping for 
$k=2$ only. For $e > 0.005$, there is no spurious damping.

\subsection{Mass Accretion Procedure}
\label{sec:accretion_procedure}
Accretion onto the protoplanet was simulated by removing material within 
a distance of $\racc=0.3\,\Rhill$. Mass is removed by means of a two-step 
procedure according to which the removal timescale, $\tacc$, is $0.03$ 
orbital periods for $\mdp < \racc/2$  and $0.09$ orbital periods in the 
outer part of the accretion region, $\racc/2< \mdp < \racc$. Notice that 
$\tacc$ is about equal to the Keplerian period around the protoplanet 
(i.e., in the circumplanetary disk) at $\mdp=\racc/2$. Tests were carried 
out to evaluate the sensitivity of the procedure to both parameters $\racc$ 
and $\tacc$. These tests indicate that, if $\racc$ is reduced by a factor 
of $1.5$, the average mass accreted during an orbit varies by less than 
$10$\%. Increasing the removal timescale by a factor of $5/3$ only affects 
the accretion rate by $5$\% \citep[see also][]{tanigawa2002}. We also checked 
whether the accretion parameters influence the orbital evolution of the 
planet. Using the same tests, we found no significant differences over a 
few hundred periods of evolution.

We do not add the removed mass to the planet mass over the course of the 
simulations. Doing so would increase the planet mass by $\sim 1\,\MJup$ 
for the simulations reported here.

\subsection{Boundary Conditions}
\label{sec:boundary_conditions}
Boundary conditions at the radial inner boundary, $r=r_{\mathrm{min}}$, 
allow outflow of material, i.e., the accretion flow toward the central 
star. The inner boundary conditions exclude inflow away from the star 
into the computational domain. Two types of boundary conditions were 
used at the outer boundary ($r=r_{\mathrm{max}}$): reflective and 
non-reflective. With the first kind, neither inflow nor outflow of material 
is permitted at the outer border, which behaves as a rigid wall. Although 
$r_{\mathrm{max}}$ is much larger than the apocenter radii of the planets 
in the simulations, wave reflection was observed at the outer border in 
this case, especially when the planet orbit was eccentric.

In order to lessen the amount of wave reflection at the outer disk edge, 
we also applied non-reflective boundary conditions, following the approach 
of \citet{godon1996,godon1997}. In these circumstances, boundary conditions 
are not directly imposed on the primitive variables (i.e, $\Sigma$ and 
$\gvec{u}$), but rather on the characteristic variables (i.e., the Riemann 
invariants of one-dimensional flows). The basic idea is to let outflowing 
(inflowing) characteristics propagate through the grid boundary. The correct 
propagation of the flow characteristics across the border depends on how 
accurately the adopted solution exterior to the grid approximates the exact 
solution.

At both disk radial boundaries, the flow is assumed to be Keplerian around 
the central star. This choice could lead to some small-amplitude wave 
excitation at $r=r_{\mathrm{max}}$, since the fluid tends to orbit about 
the center-of-mass of the system, rather than around the central star
\citep[see also][]{rnelson2000}.
The effects of such waves were checked to be unimportant as they tend to
dissipate within a short distance from the outer disk boundary

\section{Disk Eccentricity}
\label{sec:disk_eccentricity}
\subsection{Global Density Distribution}
\label{sec:gobal_sigma}
\begin{figure*}
\centering%
\resizebox{0.82\linewidth}{!}{%
\includegraphics{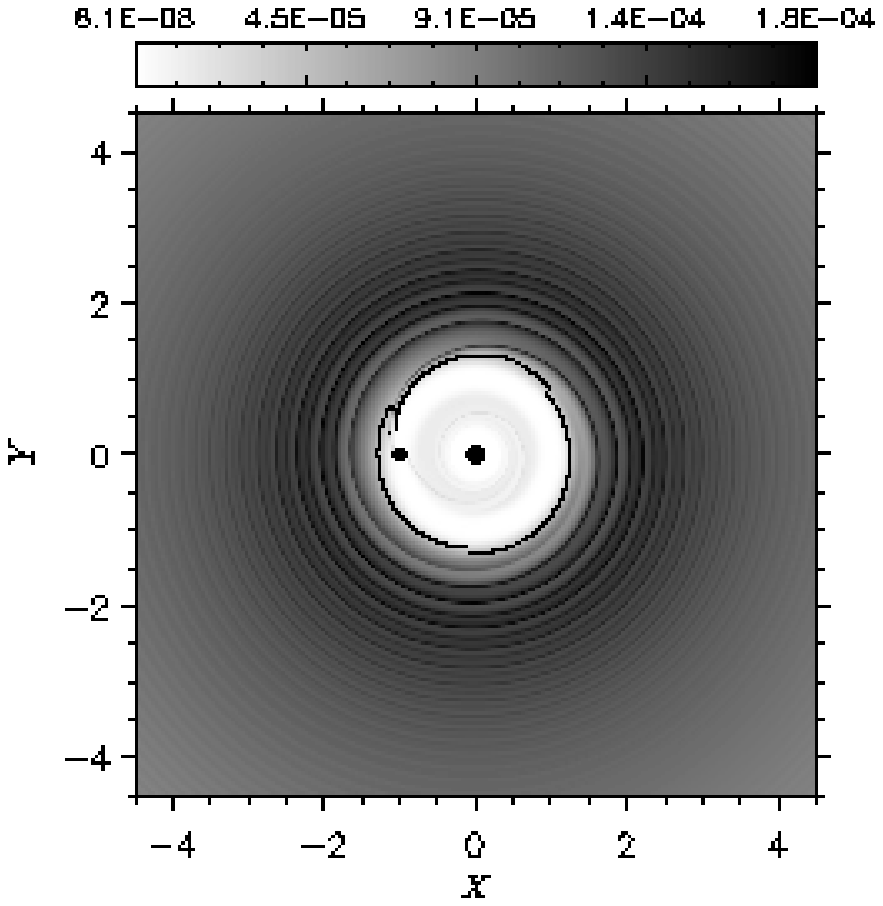}%
\includegraphics{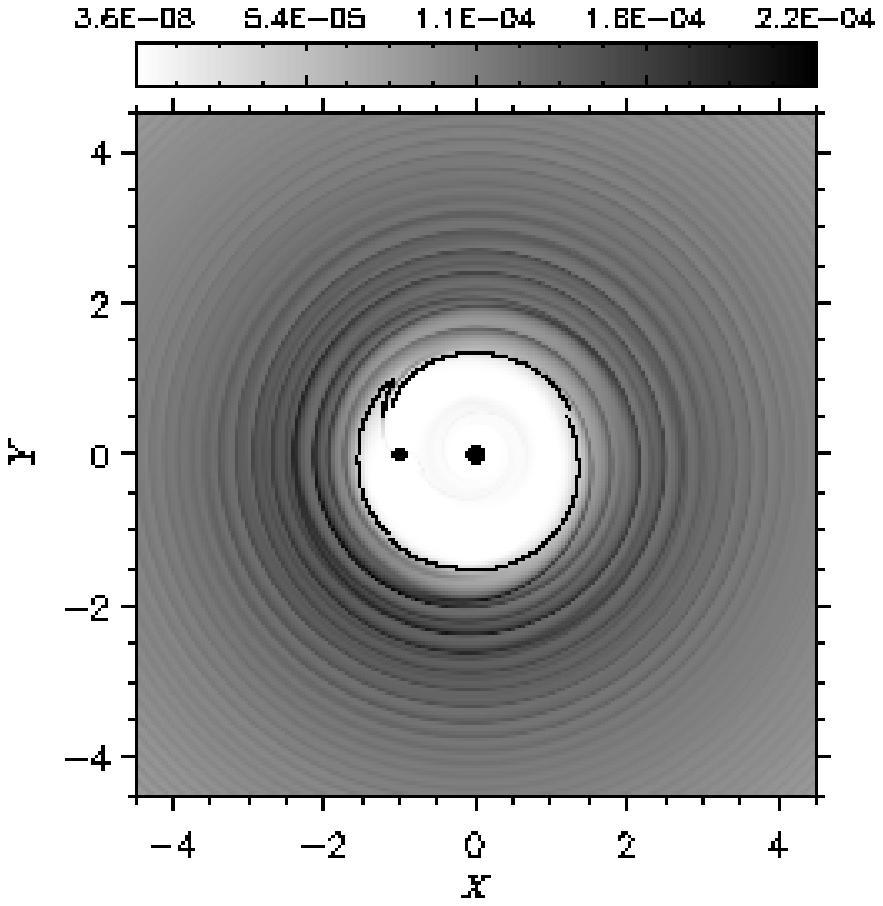}}
\resizebox{0.82\linewidth}{!}{%
\includegraphics{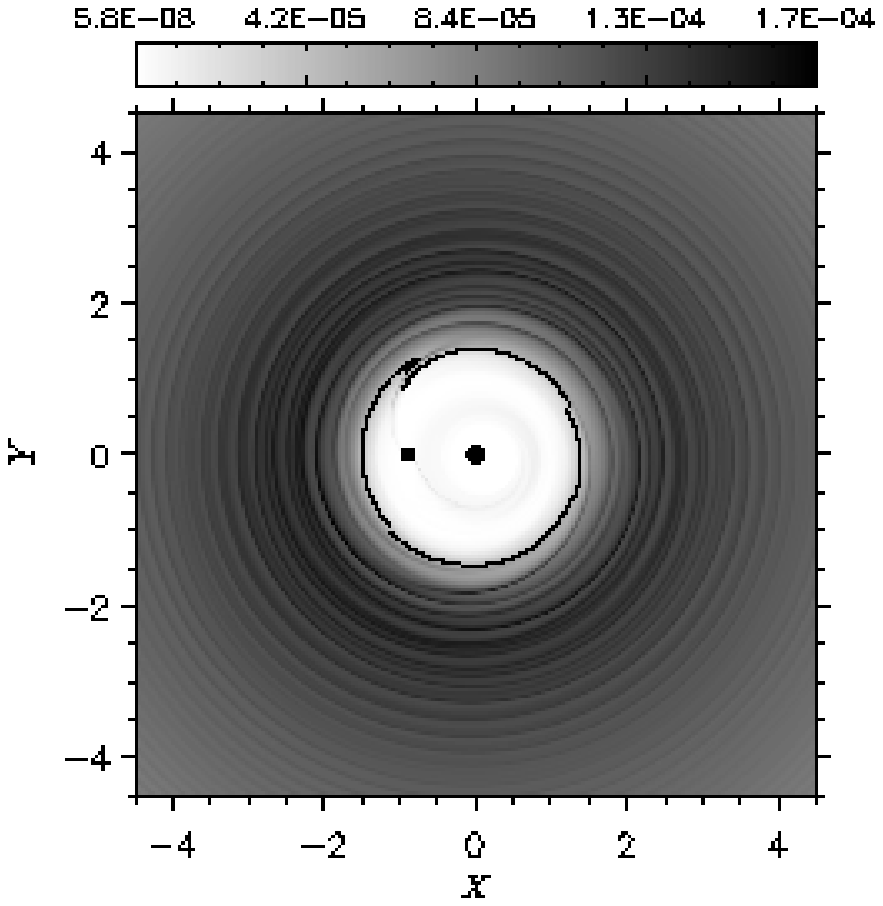}%
\includegraphics{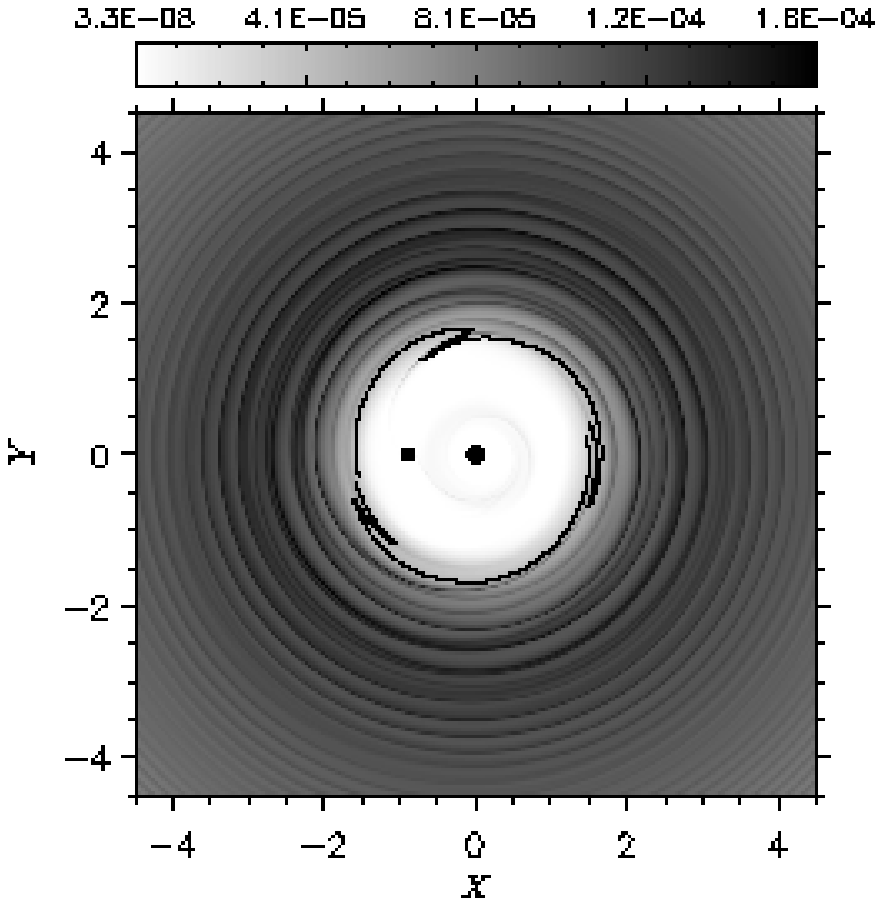}}
\resizebox{0.82\linewidth}{!}{%
\includegraphics{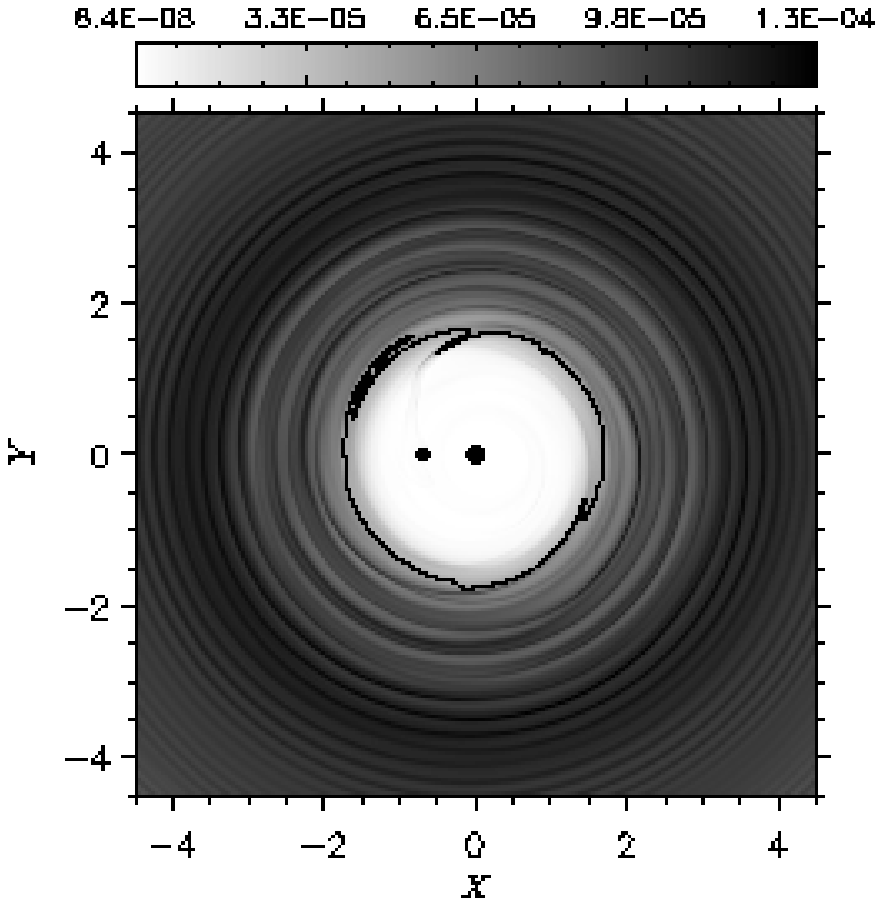}%
\includegraphics{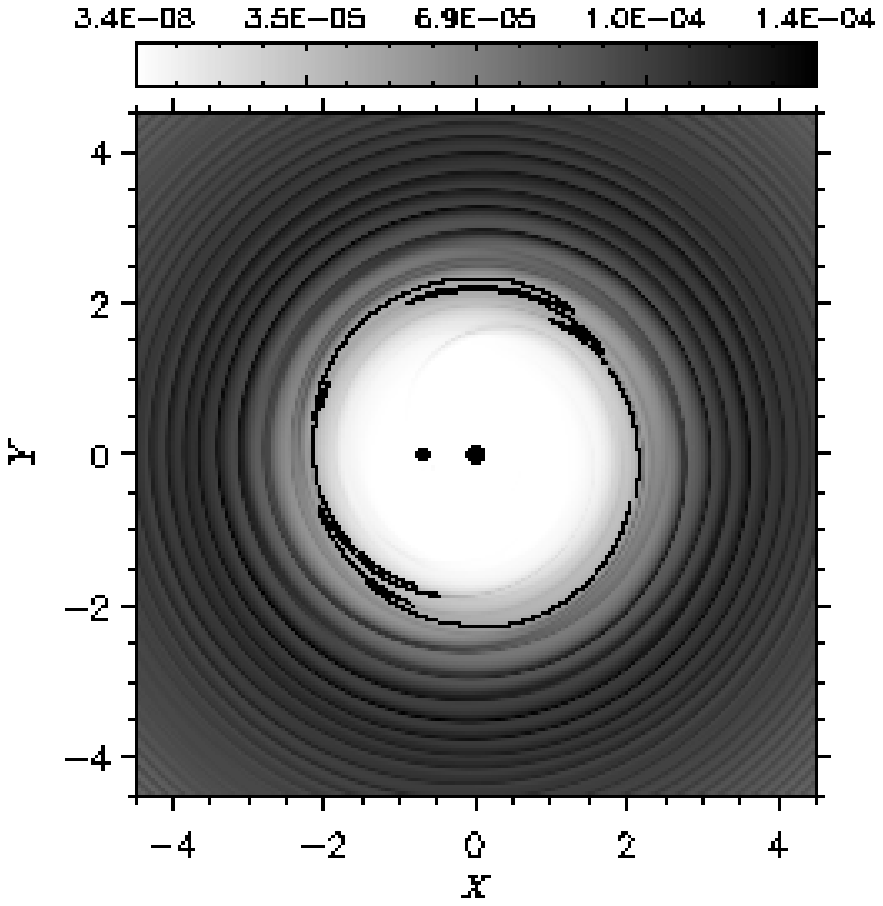}}
\caption{\small%
         Surface density in a disk containing a $1\,\MJup$ (left) and a
         $3\,\MJup$ (right) planet at $t=1000$ orbits. The axes are in 
         units of the planet's orbital semi-major axis, $a_{0}$. The grey 
         scale bars are expressed in units of $3.29\times10^5\,\sdunits$.
         From top to bottom, panels refer to the configurations with orbital 
         eccentricities $e=0$, $0.1$, and $0.3$. In each panel the planet 
         (smaller circle) is at pericenter and located at $(X,Y)=(e-1,0)$.
        }
\label{fig:globalsigma}
\end{figure*}
\begin{figure*}
\centering%
\resizebox{0.95\linewidth}{!}{%
\includegraphics{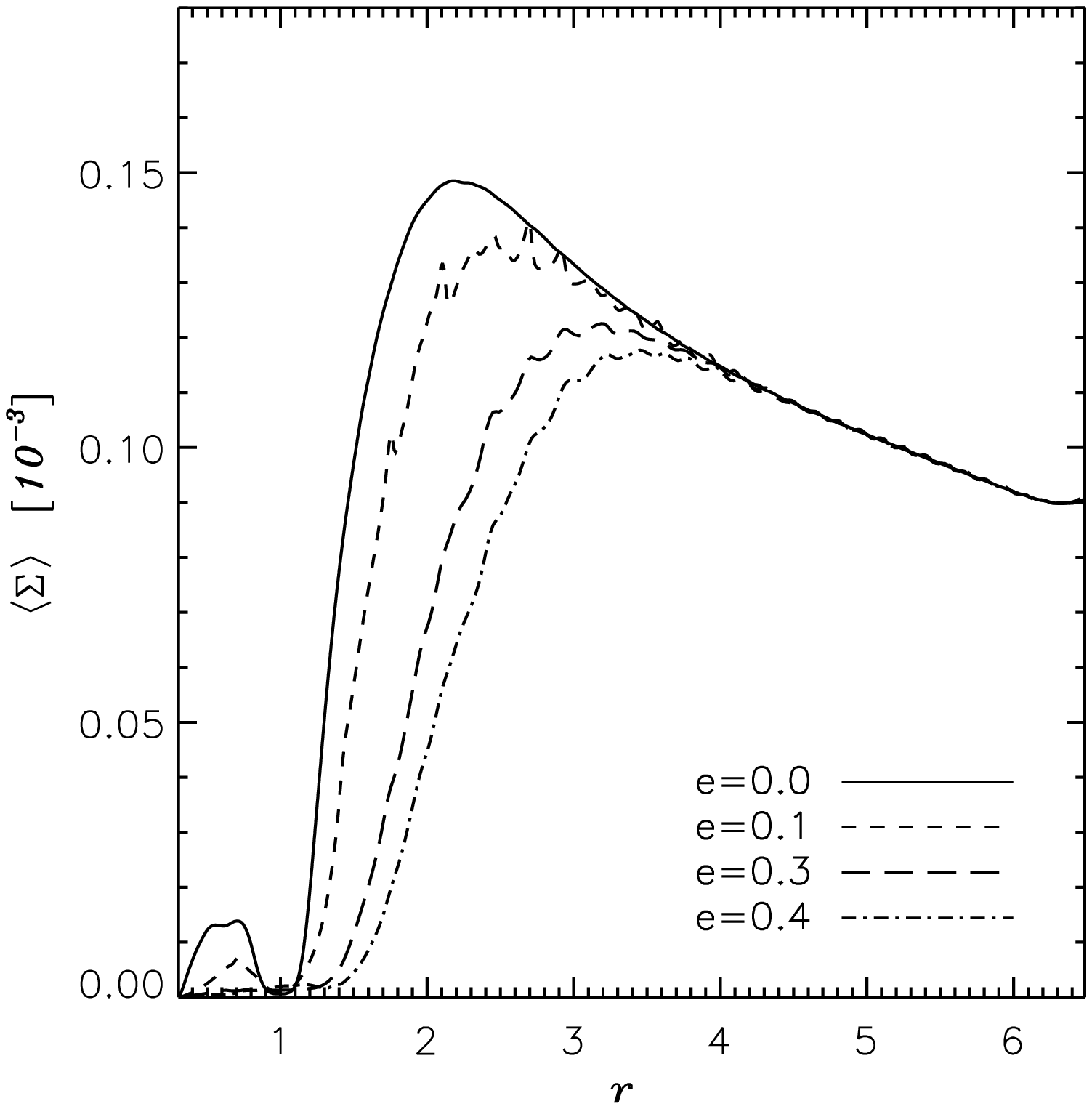}%
\includegraphics{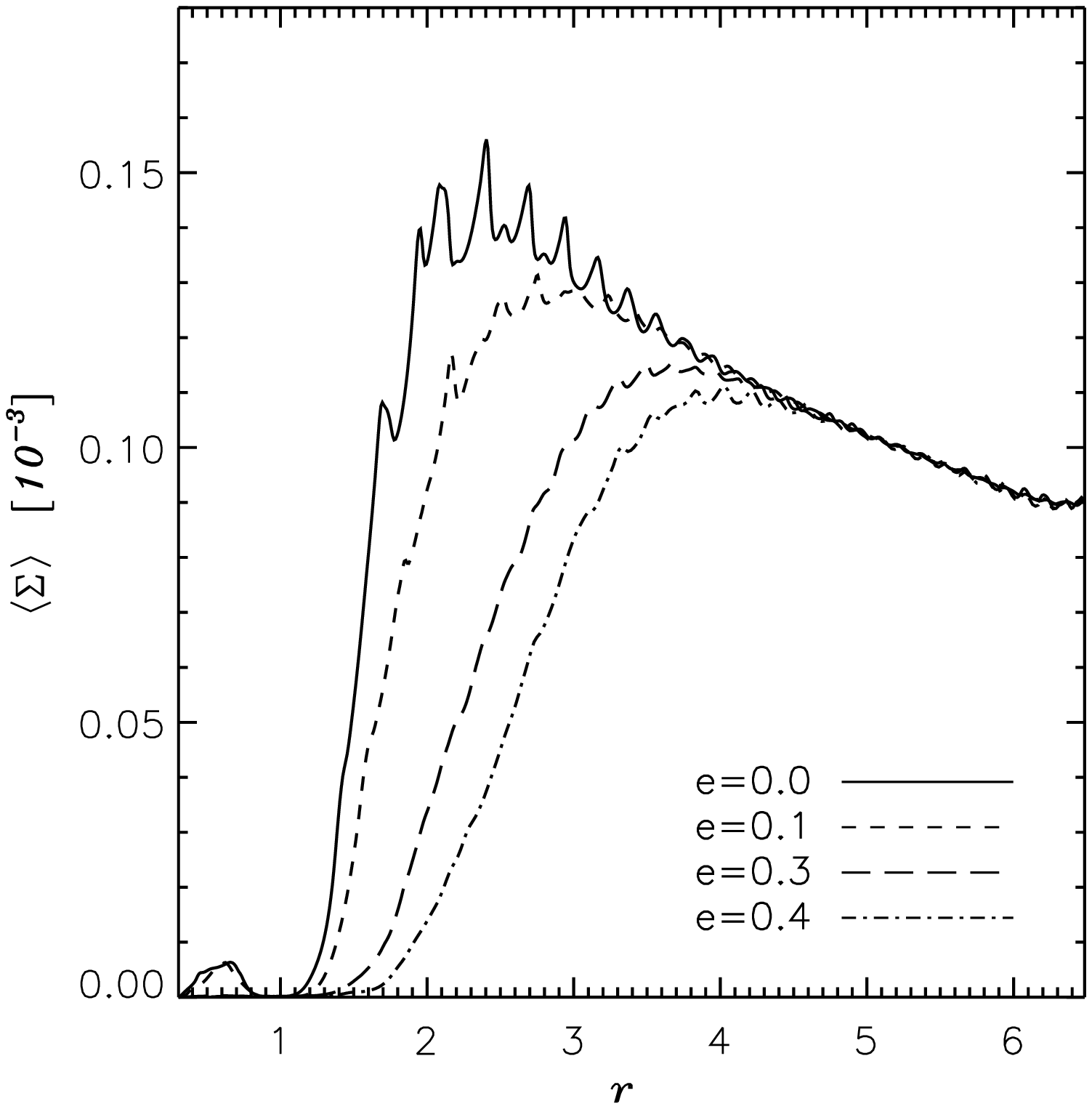}}
\caption{\small%
         Azimuthally averaged surface density in a disk containing
         a $1\,\MJup$ (left) and $3\,\MJup$ (right) planet for various
         values of orbital eccentricity: $e=0$ (solid line), $e=0.1$ 
         (short-dashed line), $e=0.3$ (long-dashed line), and $e=0.4$ 
         (dotted-dashed line). The horizontal axis is in units of the planet's 
         orbital semi-major axis $a_{0}$. For the vertical axis, 
         $\langle\Sigma\rangle=10^{-4}$ corresponds to $32.9\,\sdunits$. 
         The time is about $1000$ orbits and the planet is at pericenter. 
        }
 \label{fig:avsigma}
\end{figure*}
\begin{figure*}[t!]
\centering%
\resizebox{0.9\linewidth}{!}{%
\includegraphics{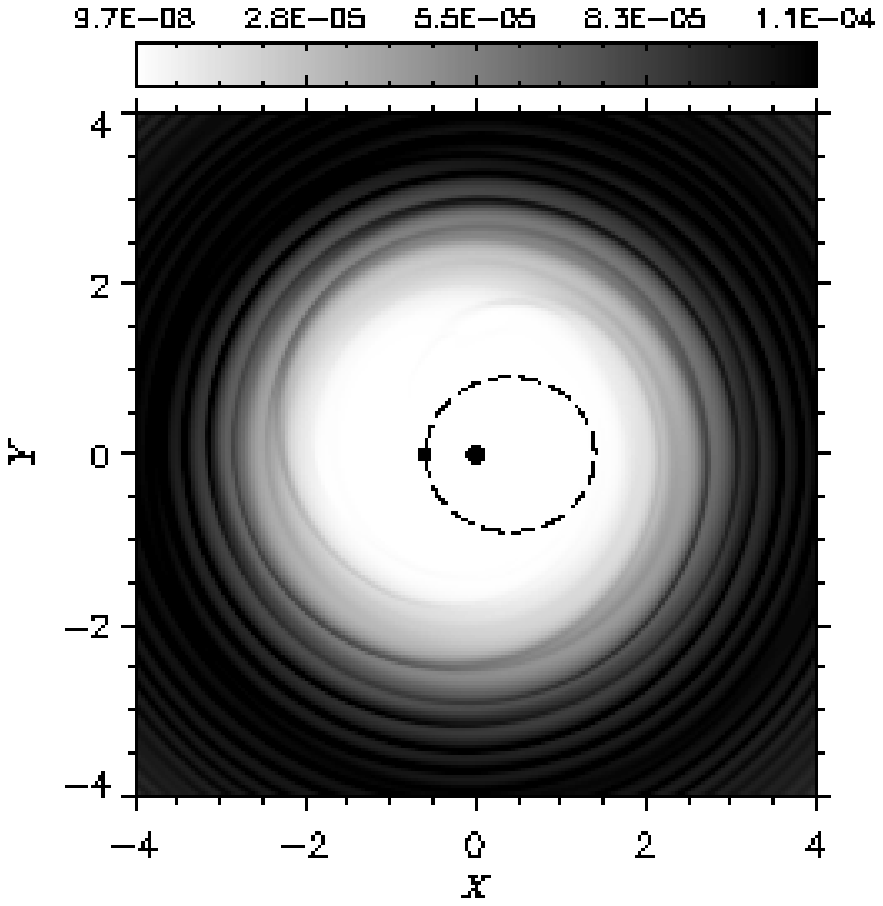}%
\includegraphics{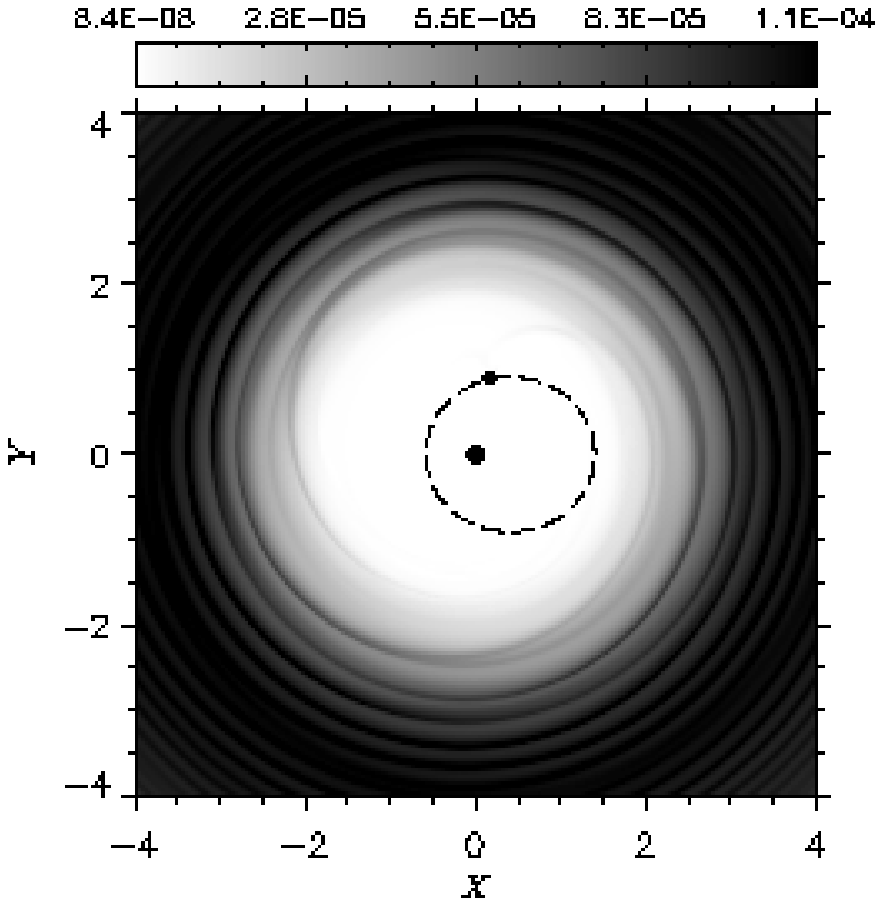}}
\resizebox{0.9\linewidth}{!}{%
\includegraphics{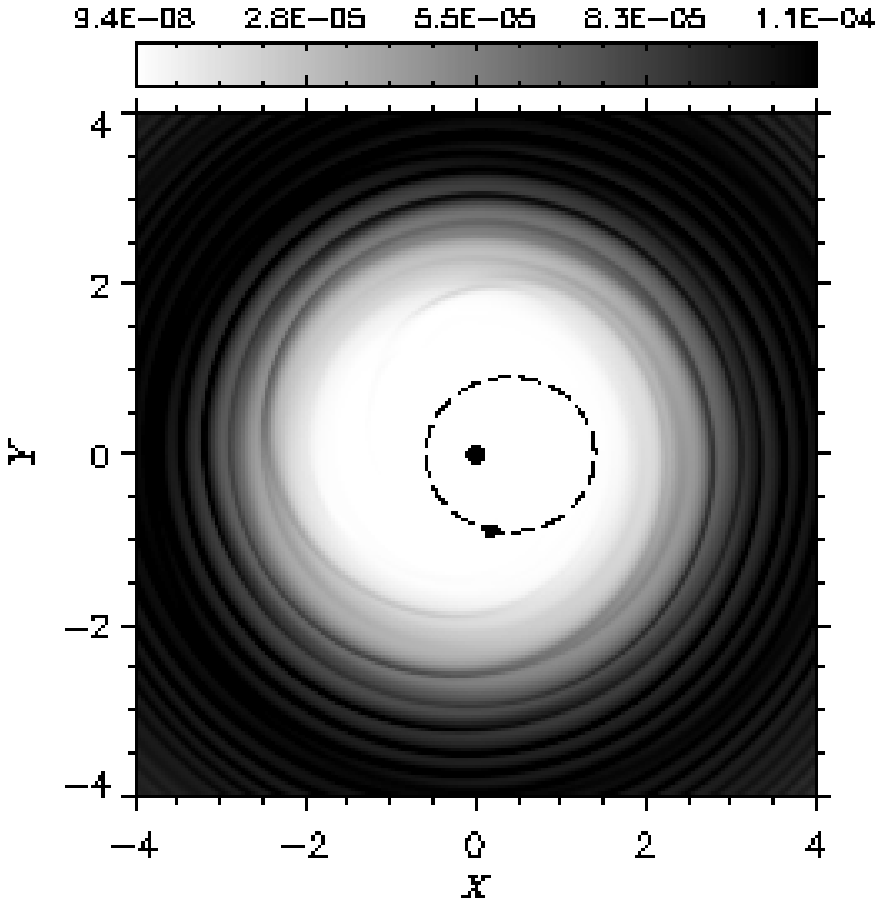}%
\includegraphics{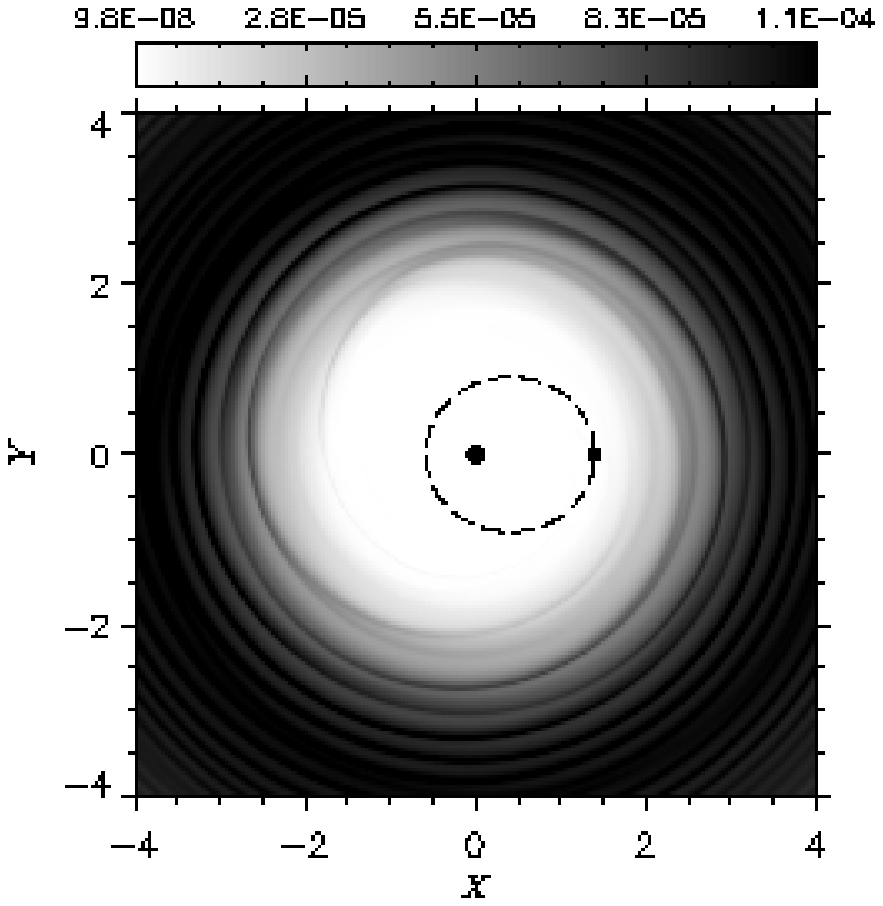}}
\caption{\small%
         Gap region around the orbit of a $3\,\MJup$ planet with fixed 
         $e=0.4$ at various orbital phases. The axes are in units of 
         the planet's orbital semi-major axis $a_{0}$. The grey scale 
         bars are expressed in units of $3.29\times10^5\,\sdunits$. The
         planet's orbit in the inertial frame is represented by a dashed 
         ellipse. The smaller solid circle indicates the position of the
         planet around the star (larger solid circle). The motion of the
         planet (and the panel sequence) is counter-clockwise.
        }
\label{fig:gapedge_inertial}
\end{figure*}

The global surface density in the disk is plotted in 
\refFgt{fig:globalsigma} for different planet's masses and orbital 
eccentricities. Left panels refer to configurations with $\Mp=1\,\MJup$,
and right panels refer to configurations with $\Mp=3\,\MJup$. The orbital 
eccentricity increases from top ($e=0$) to bottom ($e=0.3$). In all of the 
panels, the planet is at pericenter ($\rp/a=1-e$). For sufficiently long 
evolutionary times, the inner disk  (disk interior to the planet) is 
largely depleted because of the tidal gap  produced by the planet and 
because the grid does not have sufficient dynamic range to cover regions 
close to the star where an inner disk  would reside. The outer disk is 
tidally truncated at a radial distance that depends on both $\Mp$ and $e$.  
\refFgt{fig:globalsigma} shows that the size of the truncation radius 
increases with planet mass and eccentricity, analogous to the case for 
circumbinary disks \citep{pawel1994}. In addition, a wave or wake propagates 
in the outer disk. The disk truncation also be seen in \refFgt{fig:avsigma},
which shows the azimuthally averaged surface density for various cases.
The outer gap edge becomes less steep as the orbital eccentricity increases 
and does not change significantly over an orbital period, as illustrated in 
\refFgt{fig:gapedge_inertial}.

\begin{figure}
\centering%
\resizebox{1.0\linewidth}{!}{%
\includegraphics{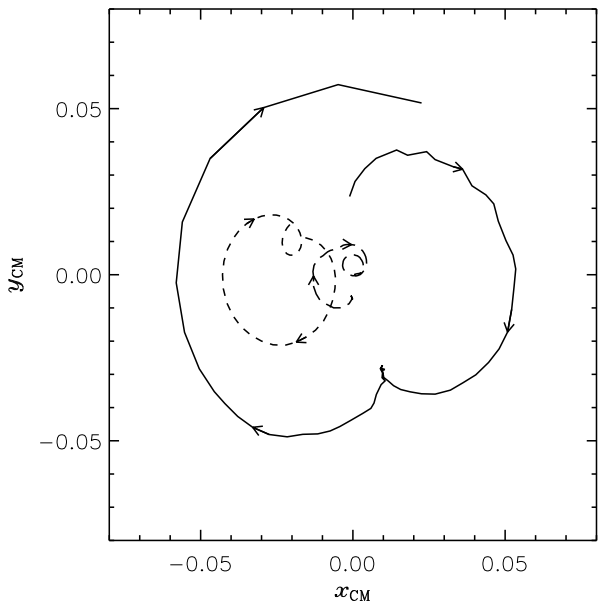}}
\caption{\small%
         Motion of the center-of-mass of the disk region $r<3\,a_{0}$
         relative to the star (located at the origin). The disk
         center-of-mass starts at the origin, but is plotted at later times
         $t\gtrsim 300$ orbits. The long-dashed and solid  lines refer to 
         the cases with $\Mp=1\,\MJup$ and $\Mp=3\,\MJup$, respectively, 
         with a fixed circular orbit planet. The short-dashed line is for 
         the model with $\Mp=3\,\MJup$ and fixed planet eccentricity $e=0.1$.
        }
\label{fig:rcmvstime}
\end{figure}

After a few hundred orbits, the disk region near the planet  becomes 
eccentric, even if the planet is on a circular orbit. This effect can be 
seen in \refFgt{fig:rcmvstime}, which illustrates the motion (relative to 
the star) of the center-of-mass of the disk interior to $r=3\,a_{0}$ for 
different planetary masses and orbital eccentricities. 

\subsection{Mode Analysis}
\label{sec:modes_analysis}
We performed a mode decomposition of the surface density distribution of 
the disk adopting an approach along the lines of \citet{lubow1991b}.
We defined a mode component at each radial location in the disk as
\begin{equation}
 \label{eq:modeMkl}
 \mathcal{M}_{(k,l)}^{(f,g)}=\frac{2}{\pi\,T\,\langle\Sigma\rangle\,%
                             (1+\delta_{l,0})}\!%
                             \int_{T}\!\int_{0}^{2\pi}\!\!%
                             \Sigma\,%
         f(k\theta)\,g(l\Omega_{\mathrm{p}}t)\,\mathrm{d}\theta\mathrm{d}t\,,
\end{equation}
where $\Omega_{\mathrm{p}}=\sqrt{G\,(\MStar+\Mp)/a^{3}_{0}}$, the angle
$\theta$ is the azimuth relative to an \textit{inertial} reference frame,
and $2\pi\langle{\Sigma}\rangle=\int_{0}^{2\pi}\!\Sigma\,\mathrm{d}\theta$ .
The time interval $T$ is a ten-orbit period interval, beginning at a 
pericenter passage of the planet. The time integration is repeated every 
interval $T$. The functions $f$ and $g$ are either sine or cosine. This 
decomposition corresponds to a Fourier transform in both azimuth and time. 
The amplitude (or strength) of the mode is 
\begin{eqnarray}
 \label{eq:modeSkl}
 S_{(k,l)}&=& \left\{%
              \left[\mathcal{M}_{(k,l)}^{(\cos,\cos)}\right]^{2}+%
              \left[\mathcal{M}_{(k,l)}^{(\cos,\sin)}\right]^{2}+\right.%
              \nonumber\\%
          & & \left.%
              \left[\mathcal{M}_{(k,l)}^{(\sin,\cos)}\right]^{2}+%
              \left[\mathcal{M}_{(k,l)}^{(\sin,\sin)}\right]^{2} 
              \right\}^{1/2}.
\end{eqnarray}

In order to obtain the strength  of a mode integrated over a radial 
interval $[r_{1},r_{2}]$, the mode components $\mathcal{M}_{(k,l)}^{(f,g)}$ 
are first averaged between $r_{1}$ and $r_{2}$ and then substituted into 
\refEqt{eq:modeSkl}. A tidally disturbed non-eccentric disk has  modes 
present that all have $k=l$. For a disk ring to be eccentric, the mode 
strength associated with the pair $(k,l)=(1,0)$, known as eccentric mode, 
must be non-zero, i.e., $S_{(1,0)}>0$. To follow the eccentricity evolution, 
we analyze $S_{(1,0)}$ as a function of the time.

\subsection{Eccentricity Growth}
\label{sec:eccentricity_growth}
\begin{figure*}
\centering%
\resizebox{1.0\linewidth}{!}{%
\includegraphics{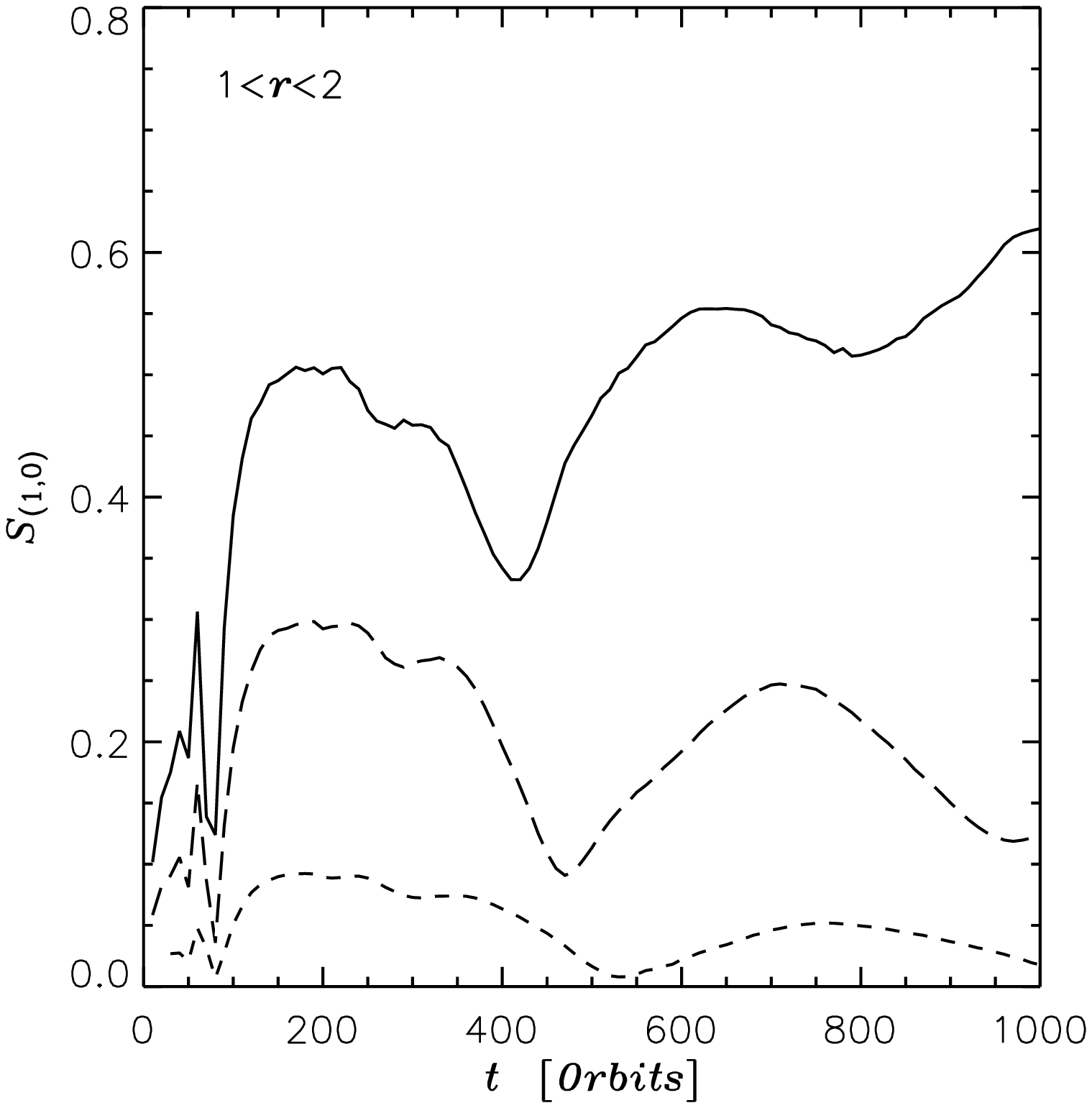}%
\includegraphics{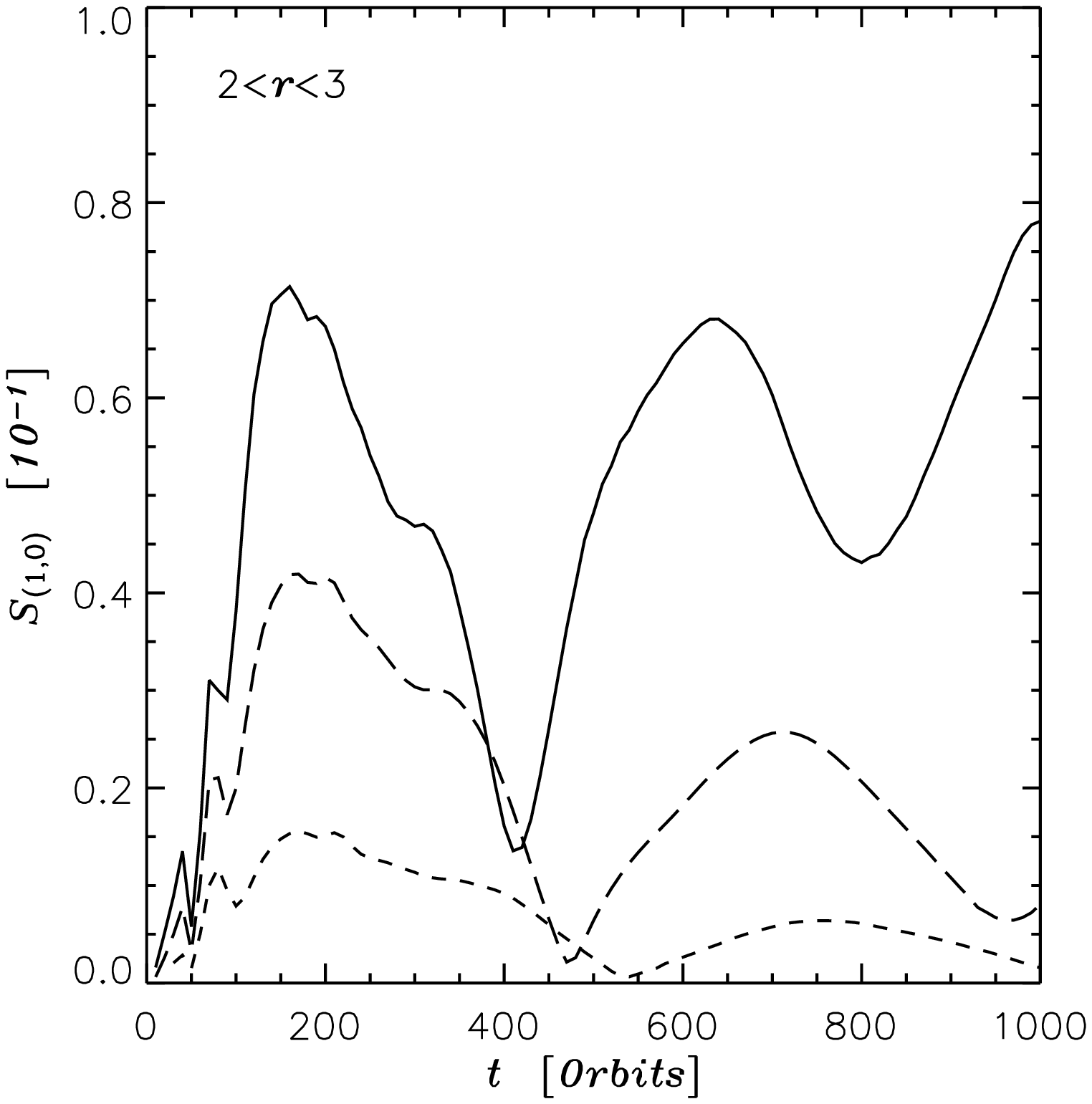}%
\includegraphics{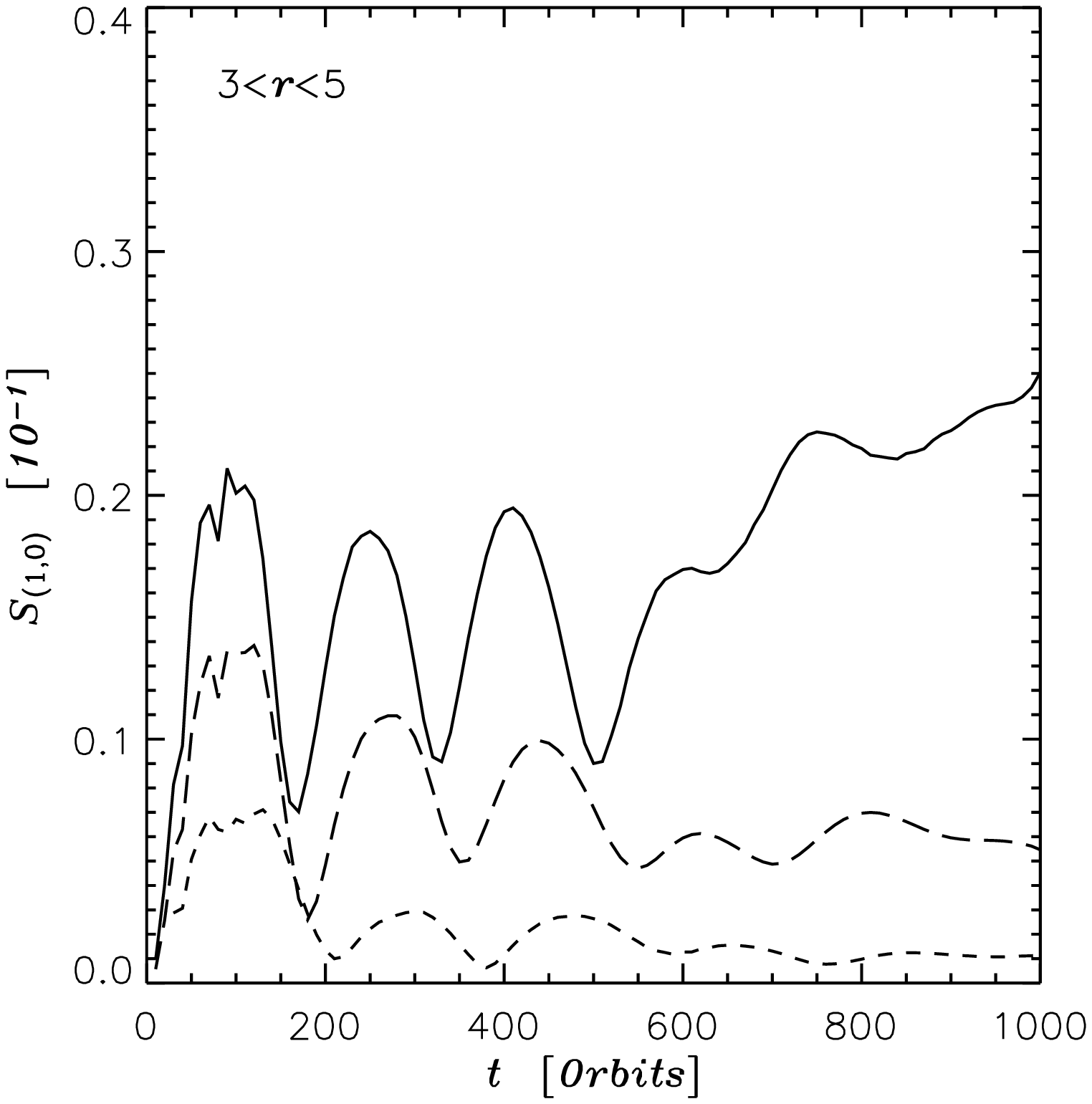}}
\caption{\small%
         Amplitudes of the eccentric mode, associated with the pair
         $(k,l)=(1,0)$ (see \refeqp{eq:modeSkl}), versus time for
         three different disk regions, as indicated in the top-left
         corner of each panel, for $1\,\MJup$ (short-dashed),
         $2\,\MJup$ (long-dashed), and $3\,\MJup$ (solid) planets.
         In all cases, the planet resides on a fixed circular orbit.
         The mode strength progressively weakens as regions farther
         from the planet's orbit are considered.
        }
\label{fig:S10_vs_t}
\end{figure*}
The amplitude of the eccentric mode, as defined in \refeqt{eq:modeSkl},
is shown in \refFgt{fig:S10_vs_t} during the evolutionary phase in which 
the planet is kept on a fixed orbit. This Figure refers to the configurations 
with $\Mp=1\,\MJup$, $2\,\MJup$, and $3\,\MJup$ planets having $e=0$.
The disk region that undergoes a substantial eccentricity growth is that
between $r\simeq a_{0}$ and $r\simeq 2\,a_{0}$ (\refFgp{fig:S10_vs_t}, left 
panel). The relatively small initial value of $S_{(1,0)}$ is likely due to 
a transient effect, as the initially circular disk adjusts to the presence 
of the planet. Farther away from the planet's orbit, the mode strength 
drastically decreases, and the disk is nearly circular (\refFgp{fig:S10_vs_t}, 
middle and right panels). The eccentricity growth proceeds very rapidly 
during the first $200$ orbits and oscillates afterwards with some 
reinforcement in the $3\,\MJup$  case.  

\begin{figure*}
\centering%
\resizebox{0.95\linewidth}{!}{%
\includegraphics{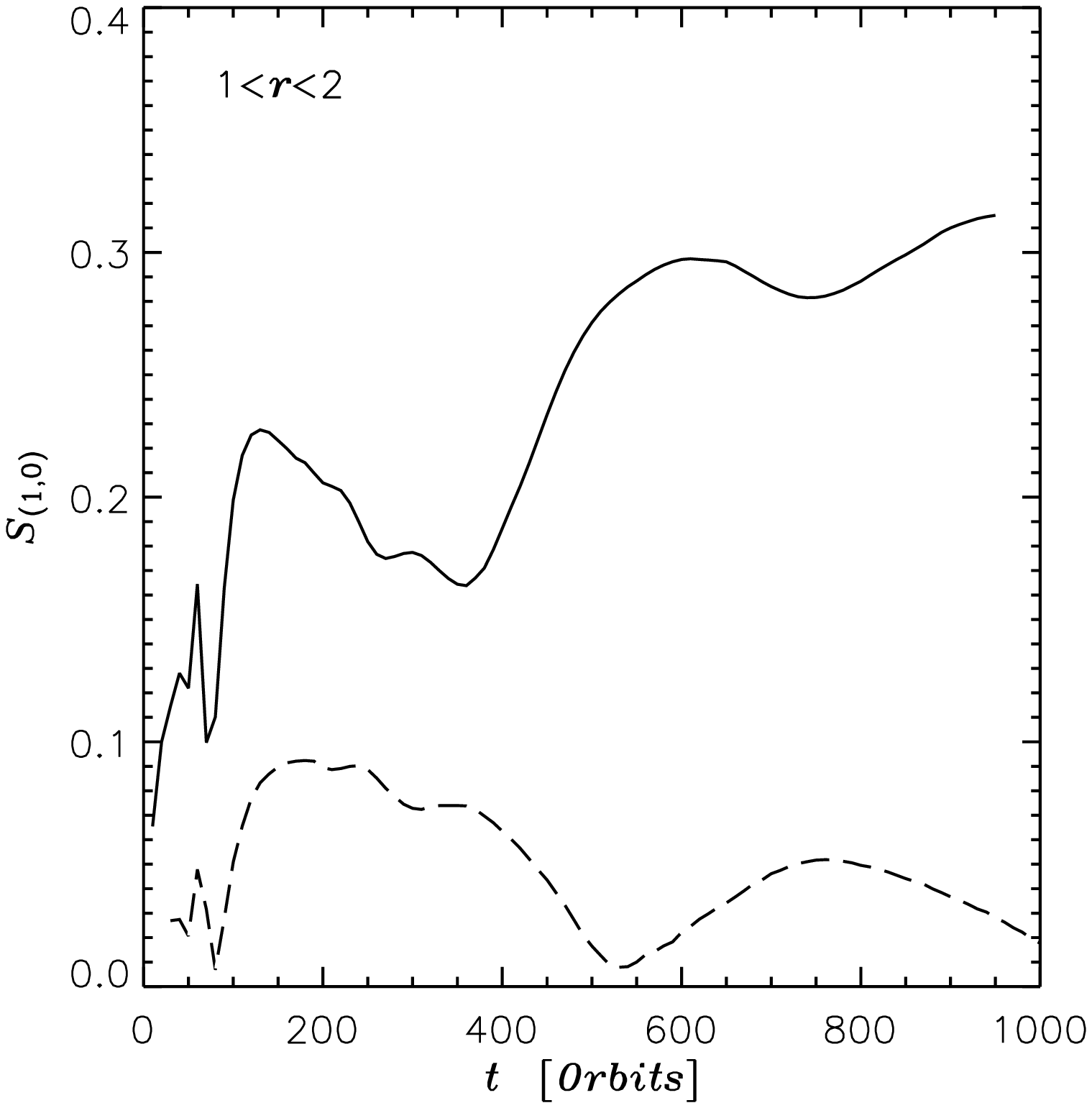}%
\includegraphics{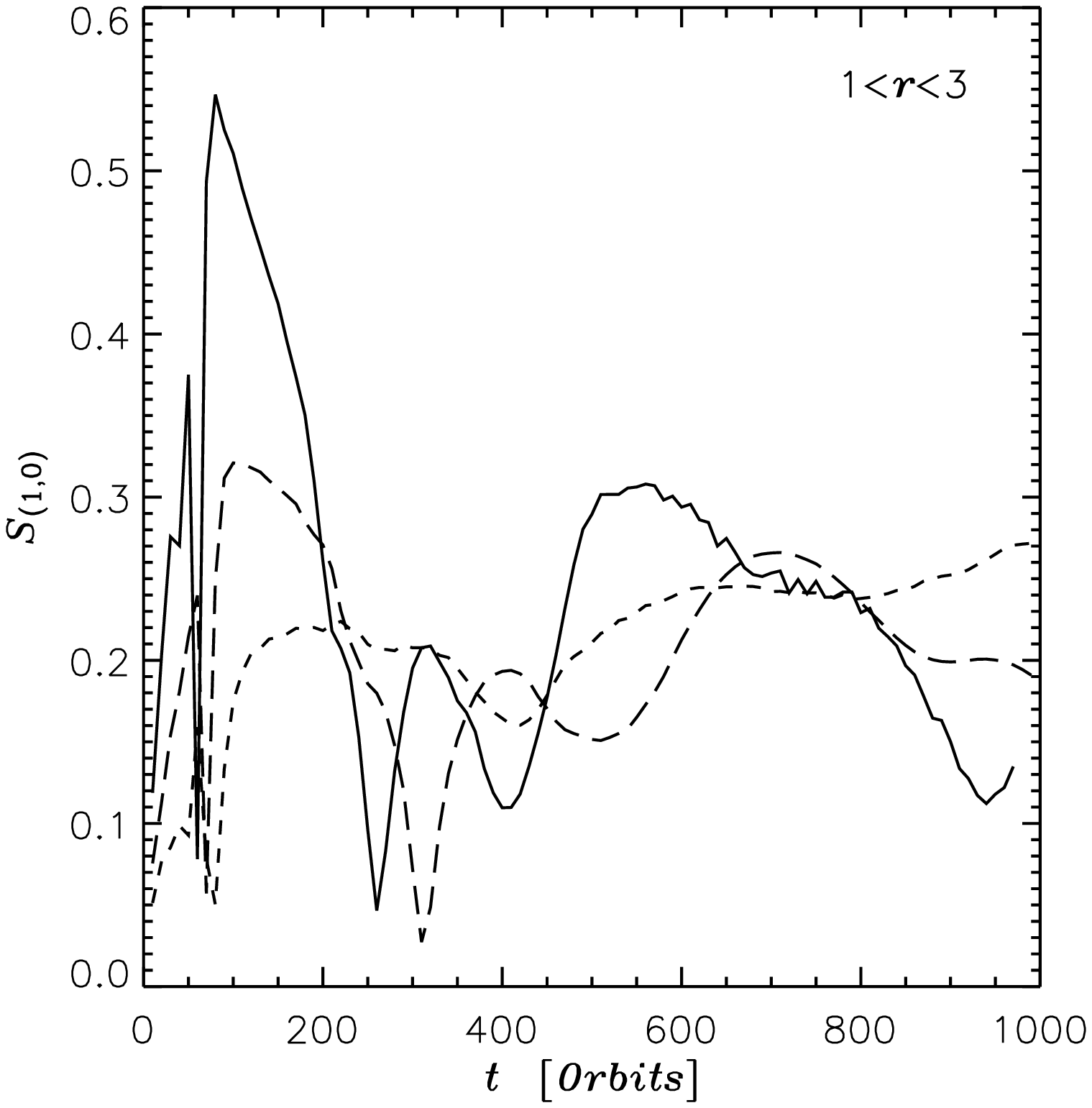}}
\caption{\small%
         Left. Mode strength $S_{(1,0)}$ as function of time for  models 
         with $\Mp=1\,\MJup$ of fixed planetary orbital eccentricity $e=0$
         (dashed line) and  $e=0.1$ (solid line).
         Right. $S_{(1,0)}$ versus time for models with $\Mp=3\,\MJup$
         of fixed planetary orbital eccentricity $e=0$ (short-dashed line),
         $e=0.1$ (long-dashed line), and $e=0.3$ (solid line).
        }
\label{fig:S10_vs_mp}
\end{figure*}
The eccentricity driven in the region $a_{0}<r<2\,a_{0}$ by the $1\,\MJup$ 
planet ($e=0$) is rather small compared to that driven by the other two 
planetary masses. However, with an initial orbital eccentricity $e=0.1$, 
a $1\,\MJup$ planet is able to sustain eccentricity growth in the disk, 
as indicated by the solid line in the left panel of \refFgt{fig:S10_vs_mp}.
In the $3\,\MJup$ case, an initial orbital eccentricity of up to $e=0.3$ 
induces a larger amplitude eccentricity perturbation only at the beginning 
of the evolution.  But the long-term evolution of $S_{(1,0)}$ is not greatly 
affected (\refFgp{fig:S10_vs_mp}).

The simulations for mode analyses were generally performed with  the grid
GS4. We also ran simulations and computed modes by using the grid system 
GS1 and found results consistent to those obtained with the higher 
resolution grid.

\subsubsection{Influence of Viscosity}
\label{sec:viscosity_influence}
The sensitivity of disk eccentricity growth to the disk kinematic viscosity  
was examined for a system containing a $3\,\MJup$ planet on an initially
circular orbit. Two additional values of $\alpha_{0}$ (i.e., $\alpha$ at 
$r=a_{0}$) were considered: $\alpha_{0}=1.2\times10^{-3}$ and 
$\alpha_{0}=1.2\times10^{-2}$, which are respectively a factor  of $3$ 
smaller and larger than the standard value $\alpha_{0}=4\times10^{-3}$. 
In the lowest viscosity model, the overall evolution of the amplitude of 
the eccentric mode closely resembles that of the model with standard 
viscosity (see \refFgp{fig:S10_vs_t}, solid line),  but was roughly $20\%$ 
larger. In the highest viscosity model, the eccentric mode strength 
reached a maximum of $0.3$ at $t=200$ orbits and then  decays. Around 
$t=1000$ orbits, $S_{(1,0)}\approx 0.1$ and continues to decline. 
The effects of viscosity were also investigated with a calculation
involving a $1\,\MJup$ planet on a circular orbit and 
$\alpha_{0}=1.2\times10^{-3}$. In this case, the mode amplitude $S_{(1,0)}$ 
is sustained at about $0.15$ for $t>200$ orbits, unlike the case of 
declining eccentricity at higher viscosity that is displayed as a 
short-dashed line in the left panel of \refFgt{fig:S10_vs_t}.
These results suggest that disks  with $\alpha$ less than a few times 
$10^{-3}$ experience sustained eccentricity, while disks with $\alpha$ in 
excess of $\approx 10^{-2}$  do not. The weakening of disk eccentricity 
with viscosity was also found in simulations by \citet{kley2006}. 

\subsection{Analytic Model}
\label{sec:analytic_model}
In the case of superhump binaries, disk precession is dominated by the 
gravitational effects of the companion which causes prograde precession 
\citep{osaki1985}. On the other hand,  pressure provides a retrograde 
contribution which is somewhat weaker \citep{lubow1992,goodchild2006}. 
In the case of a circular orbit planet, the gravitational 
contribution  to precession is expected to be weaker than the disk's 
pressure contribution. The magnitude of the pressure induced precession 
rate is $\sim (H/r)^2\,\Omega$. For the disks simulated in this paper, 
the precession timescale is then $\sim 10^3$ orbits. We estimated the 
gravitational precession due to a $3\,\MJup$ planet  for the case plotted 
in  \refFgt{fig:S10_vs_t} at $1000$ orbits  and find that this rate is 
about $10$ times smaller than the pressure precession rate. We then expect 
the precession to be pressure-dominated and retrograde, with a  timescale 
of about $10^3$ orbits. This estimate in accord with the simulation 
results for a $3\,\MJup$ planet in a circular orbit in 
\refFgt{fig:rcmvstime}. 

The group velocity for an eccentric mode is estimated by using the 
dispersion relation for an $m=1$ disturbance in a Keplerian disk perturbed 
by pressure. The group velocity is  $v_{\mathrm{g}} \sim (H/r)^2\,\Omega r$,
where we assume the radial wavenumber $| k_{r} |\sim 1/r$ and that the 
pattern speed is small compared to $\Omega$. The group velocity leads to 
eccentricity propagation timescales of order $(r/H)^2 \sim 10^3$ orbits. 
This timescale is consistent with the localization of the eccentricity over 
course of the simulation of $\sim 10^3$ orbits to within a region of order
$2\,a$, as seen in \refFgt{fig:S10_vs_mp}. Over longer timescales the 
eccentricity would spread further.

We analyze growth of eccentricity in an outer disk that is perturbed 
by a planet on a circular orbit. Disk eccentricity growth via the 3:1 
resonance for inner disks of superhump binaries occurs on a timescale 
$\sim 0.1 w / (r q^2)$ binary orbit periods, where $w$ is the radial 
width of the eccentric region \citep{lubow1991a}. We now consider the 
case for outer disks perturbed by planets. For simplicity, we assume 
that the eccentric corotational resonances in the disk are saturated 
(i.e., of zero strength). This situation is likely to hold if the disk 
eccentricity is of order $0.01$, by analogy with the case of an eccentric 
orbit planet interacting with a circular disk \citep{ogilvie2003}. 
Following the mode coupling analysis for eccentric Lindblad resonances 
involving a circular orbit planet \citep{lubow1991a}, we find that the 
disk eccentricity growth rate associated with a particular eccentric 
outer Lindblad resonance is given by
\begin{equation}
 \lambda_m = \frac{\pi F_m^2 \Omega_{\mathrm{p}} r}{24 m w}\,,
\end{equation}
where
\begin{equation}
 F_m=\frac{ 2 r u'_m - 4 r v'_m - 2 u_m- m\, u_m + 2 v_m\,(m-1)}%
          {2 r \Omega_{\mathrm{p}}},
\end{equation}
\begin{equation}
 u_m = \frac{m (\Omega - \Omega_{\mathrm{p}})\psi_m' + 2 m \Omega \psi_m /r}%
            {\Omega^2 -m^2 (\Omega-\Omega_{\mathrm{p}})^2}\,,
\end{equation}
\begin{equation}
 v_m = \frac{\Omega \psi_m' /2 + m^2 (\Omega -\Omega_{\mathrm{p}}) \psi_m /r}%
            {\Omega^2 -m^2 (\Omega-\Omega_{\mathrm{p}})^2}\,,
\end{equation}
\begin{equation}
 \psi_m = -q \, \Omega_{\mathrm{p}}^2 a^2%
       \left[\frac{1}{\pi}\int_0^{2 \pi}
       \frac{ a \cos{(m \phi)} \, \mathrm{d}\phi}%
            {\sqrt{a^2+r^2-2 r a \cos{(\phi)}}} - \frac{r}{a} \delta_{m,1} 
       \right]\,,
\end{equation}
where $w$ denotes the radial extent of the eccentric region and prime 
denotes differentiation in $r$. All quantities are evaluated at the 
location of the eccentric Lindblad resonance associated with azimuthal 
wavenumber $m$, having $\Omega=m\Omega_{\mathrm{p}}/(m+2)$ and 
$r=r_m= [(m+2)/m]^{2/3}\,a$. Quantities $u_m$ and $v_m$ are the velocity 
components in a circular disk associated with potential $\psi_m$, and 
$\delta$ is the Kronecker delta function in the indirect potential term.

The combined effects of all the resonances is determined by the local 
disk density at each resonance. The growth rate of the mass-weighted 
eccentricity is then
\begin{equation}
 \label{eq:lambdas}
 \lambda = \frac{2 \pi w}{M_{\mathrm{e}}}%
           \sum_{m=1}^{m_{\mathrm{max}}} \Sigma_m r_m \lambda_m\,,
\end{equation}
where $M_{\mathrm{e}}$ is the mass of the eccentric region and the sum is 
taken over the active eccentric Lindblad resonances, and 
$m_{\mathrm{max}} \simeq r/H$ due to torque cut-off effects \citep{gt1980}.
 
\begin{figure}[t!]
\resizebox{1.0\linewidth}{!}{%
\includegraphics{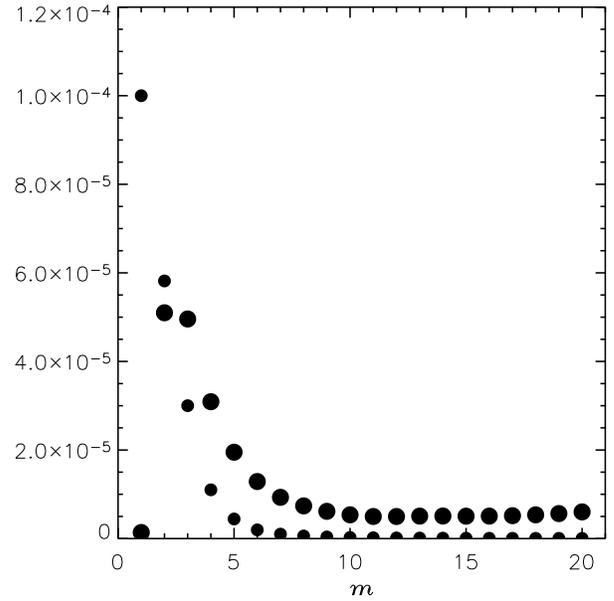}}
\caption{\small%
         Large dots: contributions to the growth rate sum for
         \refEqt{eq:lambdas} in units of $\Omega_{\mathrm{p}}$ 
         as a function of azimuthal wavenumber $m$ associated
         with an eccentric Lindblad resonance.
         Small dots: normalized azimuthally averaged surface
         density at each eccentric Lindblad resonance,  
         $10^{-4}\, \Sigma_m/\Sigma_1$, as a function of $m$.
         The plot is for a $3\,\MJup$ planet with a density 
         profile obtained from simulations at $500$ orbits.
        } 
\label{fig:lam}
\end{figure}
The contribution of each resonance in the above sum for $\lambda$
for a $3\,\MJup$ planet is shown in \refFgt{fig:lam}.  We consider
times beyond several hundred orbits, when the width of the eccentric 
region $w$ $\sim 2\,a$. \refFgt{fig:lam} shows that there is a weak 
contribution from the outermost resonance, the 1:3 resonance, corresponding 
to $m=1$. Even though the density at this resonance is the largest,  
a nearly complete cancellation occurs in $\lambda_1$, due to effects 
of the indirect term in potential $\psi_1$. If only the $m=1$ resonance 
were involved, then the eccentricity growth timescale would be very long, 
$\sim 10^5$ orbits. The growth rate contributions from regions closer 
to the planet are weakened by the lower density, but strengthened by 
the larger values of $\lambda_m$. At a time of 500 orbits, the density 
near the planet is small enough that the outermost resonances 
($2 < m \la 5$) provide most of the growth rate. At a time of 500 orbits,
the eccentricity growth timescale $1/\lambda$ is about $600$ orbits, which 
is in very rough agreement with the average growth rate implied by 
\refFgt{fig:S10_vs_t} for the innermost region, although there are 
considerable fluctuations in the simulations.\footnote{%
To compare the growth rate defined by $\lambda$ with simulations, 
it is better to adopt a similar mass-weighted eccentricity. 
This requires a slightly different definition of $S_{(1,0)}$ 
than given by \refEqt{eq:modeMkl}. When we apply that definition, 
we obtain eccentric disk evolution in the $3\,\MJup$ case that is 
similar to the $3\MJup$ results in \refFgt{fig:S10_vs_t}, except 
that the magnitude of $S_{(1,0)}$ in the innermost region is about 
a factor of 8 smaller. The average growth rate of $S_{(1,0)}$ in 
the innermost region is about the same over $1000$ orbits.} 
At earlier times, the growth rate is higher because $w$ is smaller.
We conclude that the disk eccentricity growth is possible over 
$\sim 10^3$ orbits in the case of planets because of contributions 
from several resonances that lie in the disk edge/gap region. This 
situation differs from the superhump binary case where only a single 
resonance is involved, since the disk extends relatively closer to 
the perturber in the planet case, due to its weaker tidal barrier. 

Increased viscosity affects the disk eccentricity in multiple ways.
It leads to further disk penetration of the planet's tidal barrier,
which could lead to stronger eccentricity growth. On the other hand, 
viscosity also acts to unsaturate (or strengthen) the corotation 
resonances which act to damp eccentricity. Furthermore, the viscosity 
acts to damp the non-circular motions associated with the eccentricity.

\section{Planet Orbital Evolution}
\label{sec:orbital_evolution}
At time $t=t_{\mathrm{rls}}$ (between $1000$--$1200$ orbits) the planet
was allowed to adjust its orbit in response to the gravitational forces
exerted by the surrounding disk material. We generally neglected torques 
on the planet due to gas within a distance of $0.5\,\Rhill$ from the planet.
However, analyses of the torque distribution at the release time indicate 
that torques within the Roche lobe do not dominate the orbital evolution of 
the planet. So we believe this procedure does not likely lead to major 
errors in the planet's orbital evolution. All calculations presented in 
this section employed grid GS2, except for the $3\,\MJup$ cases on 
initially circular orbits ($e_{0}=0$), which employed grid GS1. In order to 
analyze the orbital evolution of the planet after release, we calculated 
the osculating elements of the orbit each few hydrodynamics time-steps 
(approximately every $0.01$ orbits). To remove short-period oscillations, 
we computed the mean orbital elements \citep{beutler2005} by using an 
averaging period of one orbit. Throughout the paper, the  planetary orbital 
eccentricities and semi-major axes in the simulations refer to the mean 
orbital elements.

\subsection{Orbital Eccentricity}
\label{sec:orbital_eccentricity}
\begin{figure}[t!]
\centering%
\resizebox{1.0\linewidth}{!}{%
\includegraphics{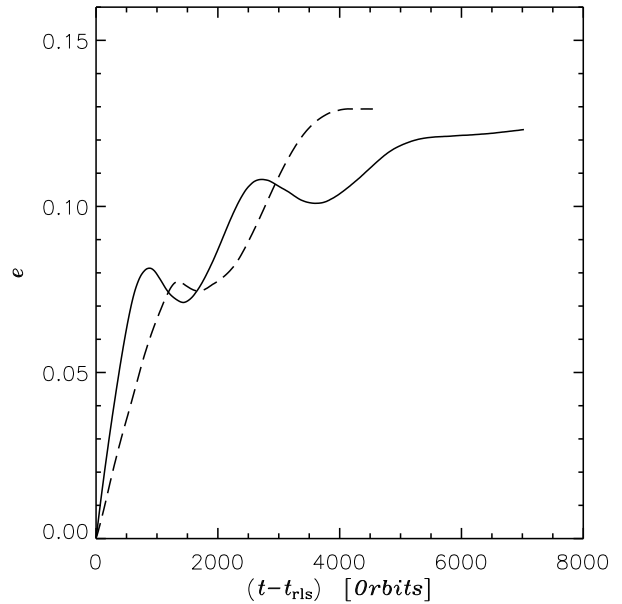}}
\caption{\small%
         Evolution of the (mean) orbital eccentricity of $2\,\MJup$ 
         (dashed line) and  $3\,\MJup$ (solid line) planets after the 
         release time, $t_{\mathrm{rls}}=1000$ orbits. 
        }
\label{fig:et_mp23}
\end{figure}
Simulations by \citet{papa2001} showed that the interaction between 
an initially circular disk and a circular orbit planet with mass 
$\gtrsim 20\,\MJup$ can lead to the growth of disk eccentricity and  
planetary orbital eccentricity. They also found that this interaction 
can be more efficient at augmenting orbital eccentricity than direct 
wave excitation at the outer 1:3 Lindblad resonance in a non-eccentric 
disk \citep[e.g.,][]{pawel1992}. We aim at determining whether a similar 
phenomenon can occur also in the Jupiter-mass range. 

\refFgt{fig:et_mp23} shows that the interaction between a planet and a 
disk leads to orbital eccentricity growth for the $2\,\MJup$ (dashed line) 
and $3\,\MJup$ (solid line) cases. During the initial growth of $e$ for 
the $3\,\MJup$ planet, the rate is 
$\dot{e}\approx 1.3\times10^{-4}\,\mathrm{orbit}^{-1}$. This value is 
$\sim 1.6$ times that exhibited by the $2\,\MJup$ planet. The eccentricity 
growth stalls  when $e\simeq 0.08$ for both planet masses. The planets may 
be experiencing some variation in their eccentricity forcing due to the 
phasing of their eccentricities relative to the disks'. 
After $1500$--$1600$ orbits from the release time, the orbital eccentricity 
starts to increase again with a  growth timescale that is comparable 
to the initial growth timescale, 
$\taue\equiv e/|\dot{e}|\approx 2.3\times10^{3}$ orbits, for both planetary 
masses. The average growth timescale is shorter than the standard Type~II 
migration (or viscous diffusion) timescale (see 
\refSecp{sec:radial_migration}).
Over the last $1000$ orbits of the simulation, the eccentricity of the 
$3\,\MJup$ planet increases very slowly, at a rate 
$\dot{e}\approx 2\times10^{-6}\,\mathrm{orbit}^{-1}$.

\begin{figure}[t!]
\centering%
\resizebox{1.0\linewidth}{!}{%
\includegraphics{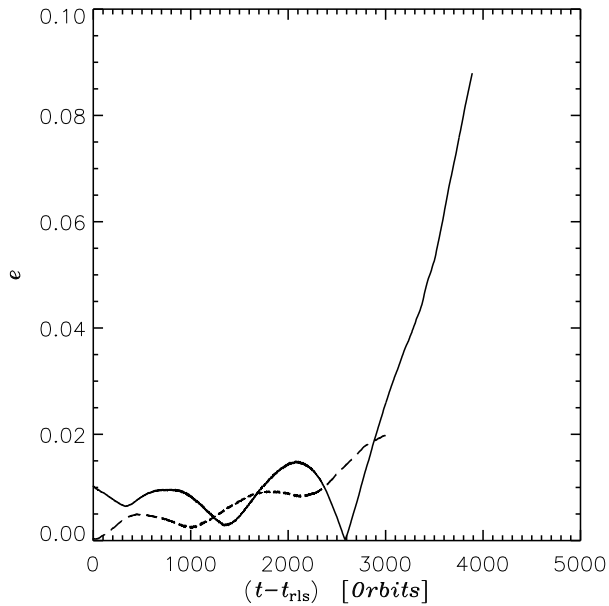}}
\caption{\small%
         Orbital eccentricity versus time of Jupiter-mass models
         with different initial orbital eccentricities: $e_{0}=0$
         (dashed line) and  $e_{0}=0.01$ (solid line). 
         The release time is $1100$ orbits.
        }
\label{fig:et_mp1vse}
\end{figure}

The model with $\Mp=1\,\MJup$ and initial zero-eccentricity 
(\refFgp{fig:et_mp1vse}, dashed line) shows a much slower orbital 
eccentricity growth, reaching $e=0.02$ after $3000$ orbits from the release 
time.  As described in \refSect{sec:eccentricity_growth}, the eccentric 
perturbation induced by the planet on the disk is also rather weak compared 
to that excited by the $2\,\MJup$ planet. At the average growth rate 
$\dot{e}\approx 7\times10^{-6}\,\mathrm{orbit}^{-1}$, it would take on the 
order of the viscous diffusion timescale to reach $e\approx 0.1$.

In order to evaluate to the extent of orbital eccentricity growth, 
we used configurations with fixed non-zero planet  eccentricities 
prior to release, $e_{0}$. \refFgt{fig:et_mp1vse} also shows the 
orbital eccentricity evolution of a $\Mp=1\,\MJup$ planet with 
$e_{0}=0.01$. After release, $e$ oscillates about the initial value. 
The oscillation grows in amplitude and, during one of these cycles, 
$e$ increases from $0$ to $0.09$ within  about $1300$ orbits. In this 
case, $\taue$ is of order the viscous diffusion timescale. We simulated 
several models with $e_{0} \ge 0.1$ (not plotted) and found that there 
was generally a reduction of the orbital eccentricity, with some 
exceptions though. For example, in a model with a $1\,\MJup$ planet 
and  $e_{0}=0.1$, $e$ underwent small amplitude oscillations about 
the initial value, with periods of a few hundred orbits. This occurrence 
may be related to the relatively large eccentricity driven in the outer 
disk. Some models with $e_{0}\ge 0.2$ showed a rate of change of $e$ 
that diminishes in time. In these cases the evolution was generally 
monitored for less than $1000$ orbits. Longer time coverage simulations 
are required to assess the long-term behavior of these configurations.

\subsection{Radial Migration}
\label{sec:radial_migration}
\begin{figure}[t!]
\centering%
\resizebox{1.0\linewidth}{!}{%
\includegraphics{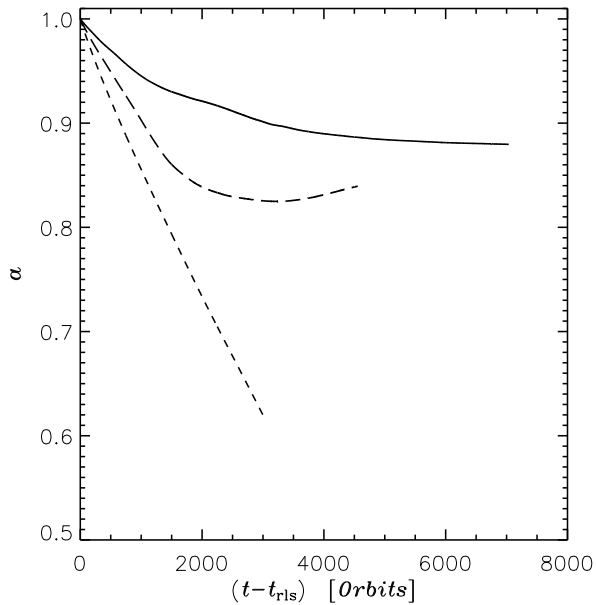}}
\caption{\small%
         Evolution of the semi-major axis of planets: $\Mp=1\,\MJup$ 
         (short-dashed line),  $\Mp=2\,\MJup$ (long-dashed line), and
         $\Mp=3\,\MJup$ (solid line). The release time is about $1000$ 
         orbits. The initial eccentricity $e_0=0$ in the $3$ cases. 
         The change in migration rates for the $2\,\MJup$ and $3\,\MJup$
         cases at later times is related to their increased orbital 
         eccentricity (see \refFgt{fig:et_mp23}).
        }
\label{fig:at_mp123}
\end{figure}
We describe here some results on the migration of eccentric orbit 
planets. We plan to explore this issue further in a future paper. 
Radial migration of planets in the mass range considered in this 
study is expected to be in the standard Type~II regime, which is 
characterized by an orbital decay timescale 
$\taum\equiv a_{0}/|\dot{a}|=2\,a^{2}_{0}/(3\,\nu)=\taum^{\mathrm{II}}$ 
\citep{ward1997}.
The Type~II migration rate depends only on the viscous timescale of
the disk near the location of the planet and is independent of the disk
density, provided the disk is locally more massive than the planet.
Type~II migration is based on the assumption that the gap, which separates the inner and outer disks, is devoid
of material. In this case, 
the planet torques approximately balance the viscous torques at the 
gap edges.

\refFgt{fig:at_mp123} plots the evolution of the semi-major axes, after 
the release time, of models with three planetary masses: $\Mp=1\,\MJup$ 
(short-dashed line), $\Mp=2\,\MJup$ (long-dashed line), and
$\Mp=3\,\MJup$ (solid line). The Figure shows that the initial migration 
rate depends on the planet's mass, which is inconsistent with the Type~II
prediction. We may expect some dependence of the migration rate on planet 
mass, because ratio of planet to disk mass is non-zero 
\citep{syer1995,ivanov1999}. In order to explore this result further, we 
have used a one-dimensional disk evolution code, along the lines of 
\citet{lin1986}. 
We used the torque density per unit mass given in equation~(4) of 
\citet{lubow2006}. We checked that the results are insensitive
to the details of the torque density, provided that it is large enough
to produce a gap. Increasing the torque density everywhere by a factor
of $2$ produced a small change in the migration rate (less than $1\%$). 
We adopted the same disk and planet parameters as in the 
two-dimensional simulations with zero planet eccentricity. In short, we 
find that the largest contributing factor to this non-Type~II behavior is 
the lack of a substantial inner disk in the two-dimensional calculations.
There is also some effect due to the non-zero planet-to-disk mass ratio. 

\begin{figure}[t!]
\centering%
\resizebox{1.0\linewidth}{!}{%
\includegraphics{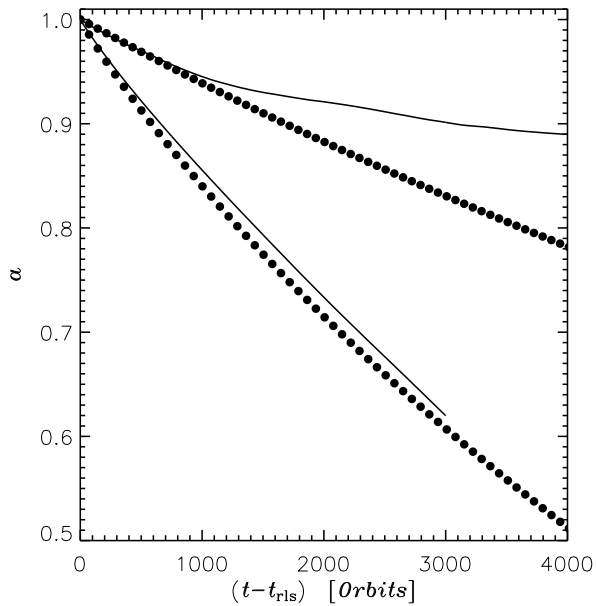}}
\caption{\small%
         Orbital migration of a $1\,\MJup$ and a $3\,\MJup$ planet 
         according to one-dimensional simulations (dots) and two-dimensional
         simulations (lines) that use the same disk and planet parameters.
         The two upper curves are for the $3\,\MJup$ planet. The 
         one-dimensional simulations have an initially depleted inner disk,
         whose density distribution matches that of the two-dimensional 
         simulations at planet release. Note the good agreement between 
         one-dimensional and two-dimensional simulations for the $1\,\MJup$ 
         planet. The agreement is also very good for the $3\,\MJup$ planet, 
         as long as its eccentricity is smaller than about $0.1$ (see 
         \refFgp{fig:et_mp23}). 
         Planet orbits are assumed to always be circular in the 
         one-dimensional models.
        }
\label{fig:at_1dvs2d}
\end{figure}
A comparison of orbital migration between one-di\-men\-sional and 
two-dimensional
models is shown in \refFgt{fig:at_1dvs2d}. In this comparison we used the
azimuthal averaged surface density distributions in \refFgt{fig:avsigma} 
as initial conditions for the one-dimensional models. As seen in 
\refFgt{fig:at_1dvs2d}, the one-dimensional (zero eccentricity) migration 
rates, for a very low density inner disk, agree well with the two-dimensional 
rates at early times after release, while the planet eccentricity is small. 
For undepleted initial inner disks, we find that one-dimensional
models have about the same migration rates for these two planet masses.
It is possible that the two-dimensional simulations have an inner boundary 
$r_{\mathrm{min}}$ that is too large to resolve the inner disk. 
More complete zone coverage of the inner region in the two-dimensional 
calculations might reveal an inner disk that acts to make the migration rate 
less dependent on mass, as indicated by the one-dimensional simulations. 
In spite of these possible limitations of our two-dimensional simulations, 
we describe below some interesting aspects of the migration of eccentric 
orbit planets in two-dimensional disks.

As a planet's orbital eccentricity grows toward values of about $0.08$,
the rate of migration slows significantly (see \refFgp{fig:at_mp123}).
Over the last $1000$ orbital periods of the calculated evolution, the 
$3\,\MJup$ planet exhibits a migration speed
$\dot{a}\approx -2\times10^{-6}\,a_{0}$ per orbit, with a tendency towards
further reduction. This migration rate is about a factor of $30$ smaller 
than the rate at release time. The $2\,\MJup$ shows an even more drastic 
reduction of the migration rate that actually reverses and becomes positive 
around $t-t_{\mathrm{rls}}\approx 4200$ orbits. The outward migration speed 
is $\dot{a}\approx 1\times10^{-5}\,a_{0}$ per orbit at the end of the 
simulation. The migration speed of the $1\,\MJup$ planet ($e_{0}=0$) 
is much more constant over the course of the simulation as its orbital 
eccentricity remains small ($e\lesssim 0.02$). 
The migration of a $1\,\MJup$ with $e_{0}=0.01$ proceeds as indicated by 
the short-dashed line in \refFgt{fig:at_mp123}. Over the last $\approx 1000$ 
orbits of evolution, however, the migration rate undergoes a decrease by a 
factor $\approx 2$. During that time, the eccentricity grows to $0.09$ 
(see solid line in \refFgp{fig:et_mp1vse}).

\begin{figure*}
\centering%
\resizebox{0.95\linewidth}{!}{%
\includegraphics{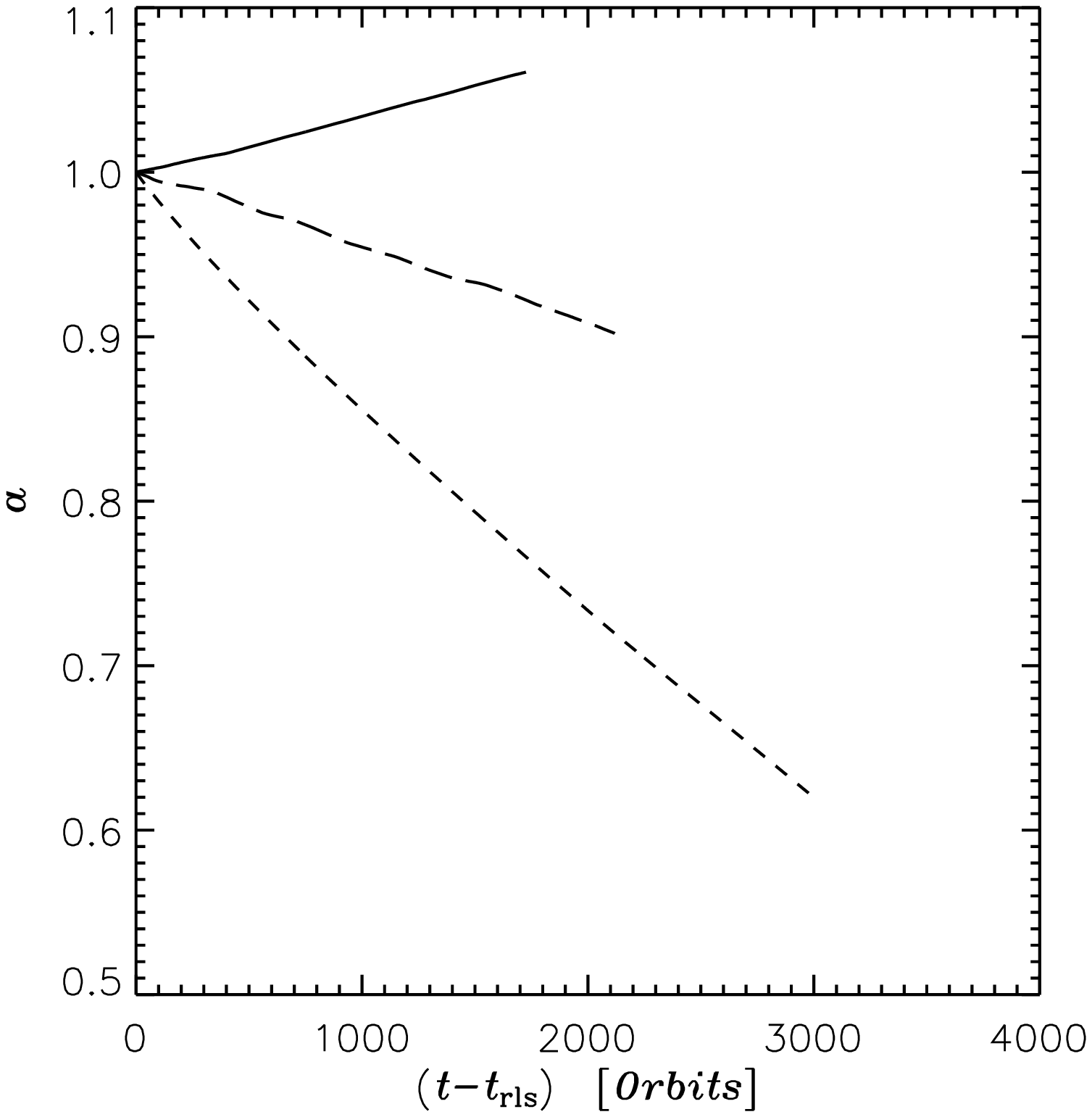}%
\includegraphics{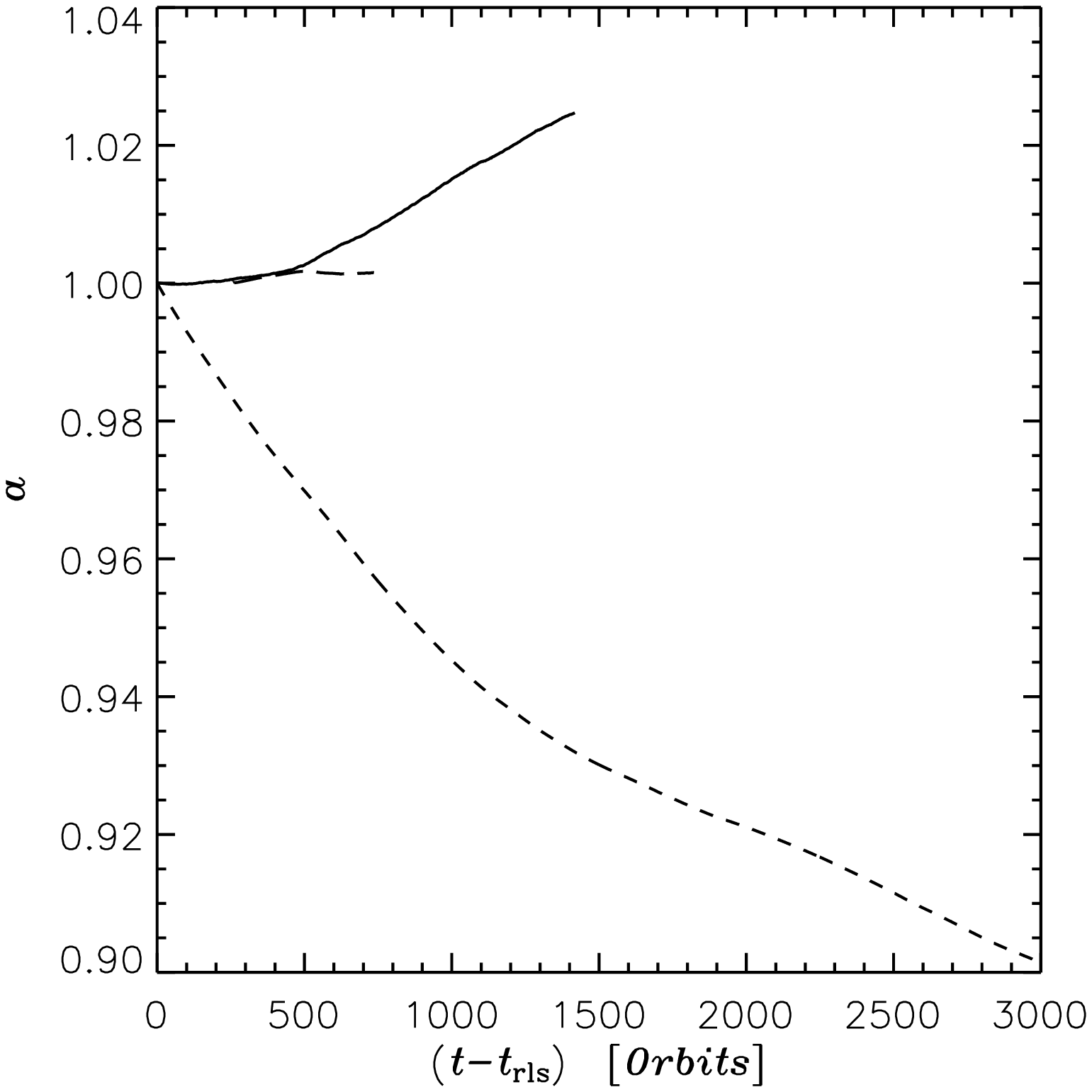}}
\caption{\small%
         Left. Semi-major axis evolution of  $1\,\MJup$ planets for three 
         values of the orbital eccentricity at release: $e_{0}=0$ 
         (short-dashed line), $e_{0}=0.1$ (long-dashed line), and 
         $e_{0}=0.2$ (solid line).
         Right. Same as the left panel but for  $3\,\MJup$ planets with
          orbital eccentricities at release: $e_{0}=0$ (short-dashed line),
         $e_{0}=0.2$ (long-dashed line), and $e_{0}=0.3$ (solid line).
         The release time is between $1000$ and $1200$ orbits.
         Our standard disk parameters were used, including 
         $\alpha = 4\times 10^{-3}$.
        }
\label{fig:at_mpvse}
\end{figure*}
The effect of non-zero orbital eccentricity on planet migration can also 
be seen in \refFgt{fig:at_mpvse}, where the evolution of the semi-major 
axis is plotted for simulations with $\Mp=1\,\MJup$ and $\Mp=3\,\MJup$ 
and different initial orbital eccentricities. There is a clear trend 
towards slower migration rates for larger orbital eccentricities. In 
particular, when $e_{0} \gtrsim 0.2$, the direction of migration is 
reversed. In all the calculations that show outward migration, the  
angular momentum of the eccentric orbit planet increases in time.

We have conducted some preliminary investigations on the cause of this 
outward torque. One possibility is that it is due to the outer disk, as
suggested in  \citet{papa2002}. When the planet eccentricity is large
enough, the angular motion of the planet at apocenter can be slower
than that of the inner parts of the outer disk, resulting in form of
dynamical friction that increases the angular momentum of the planet.
For the situation we wish to consider, it is not clear how the outer
disk inner edge is maintained, when this model is applied. This disk
material loses angular momentum from viscous torques and gains angular
momentum from the planet, in the usual torque balance for a gap. The 
latter implies that the planet should lose, rather than gain, angular 
momentum.

A preliminary analysis suggests that the outward torque may arise in the 
coorbital region. This region is supplied by material that flows from the 
outer disk across the gap, as discussed in \refSect{sec:planet_accretion}. 
In any case, further analysis is required to understand this situation.

We conducted a convergence test on the model with a $1\,\MJup$ planet
and $e_{0}=0.2$, which exhibits outward migration. The test involved
a comparison of the migration rates obtained from the grid system GS2 
(see \refTab{tbl:grids}) to those obtained from a grid system whose 
linear resolution was a factor $1.3$ larger everywhere (in both directions 
on each grid level).
However, since computing resources were only available to run the higher
resolution simulation for about $700$ orbits, we used the Gauss perturbation 
equations \citep[e.g.,][]{beutler2005} to compute $\dot{a}$ resulting from 
the disk's gravitational forces, while keeping the planet's orbit fixed. 
The result of the test is that the migration rates, averaged from $200$ to 
$700$ orbits, differed by only $7$\% at the two resolution levels.
As a check on our use of the Gauss equation, we also compared the migration 
rate determined from the Gauss equation, averaged over the last $100$ orbits 
before release, to the initial $\dot{a}$ after release, evaluated by 
integrating the equations of motion of the planet (see the left panel 
of \refFgt{fig:at_mpvse}). The two rates, both determined on grid system GS2, 
differed by less than $3\%$.
\subsection{Effects of Viscosity}
\label{sec:viscosity_effects}
\begin{figure*}[t!]
\centering%
\resizebox{0.9\linewidth}{!}{%
\includegraphics{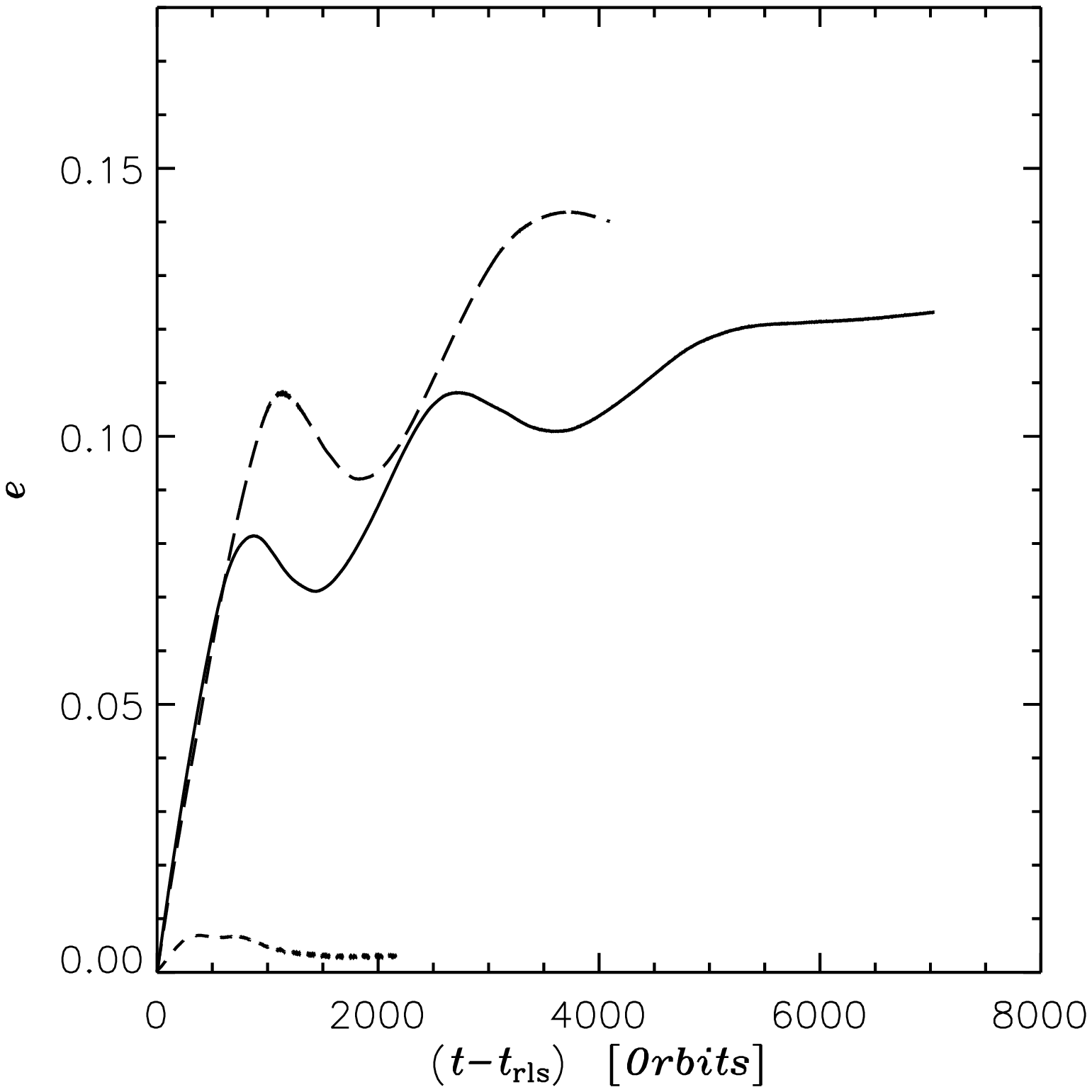}%
\includegraphics{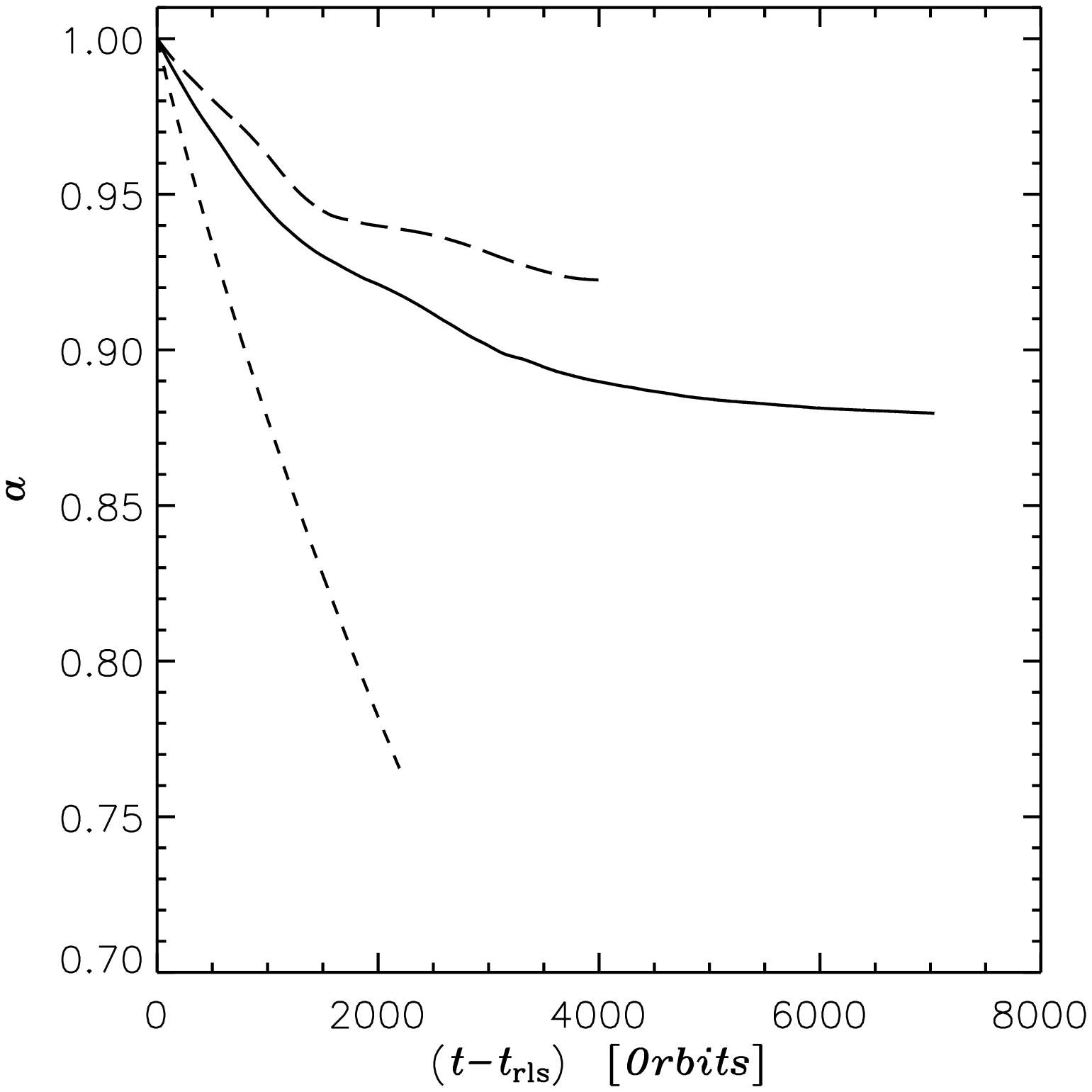}}
\resizebox{0.9\linewidth}{!}{%
\includegraphics{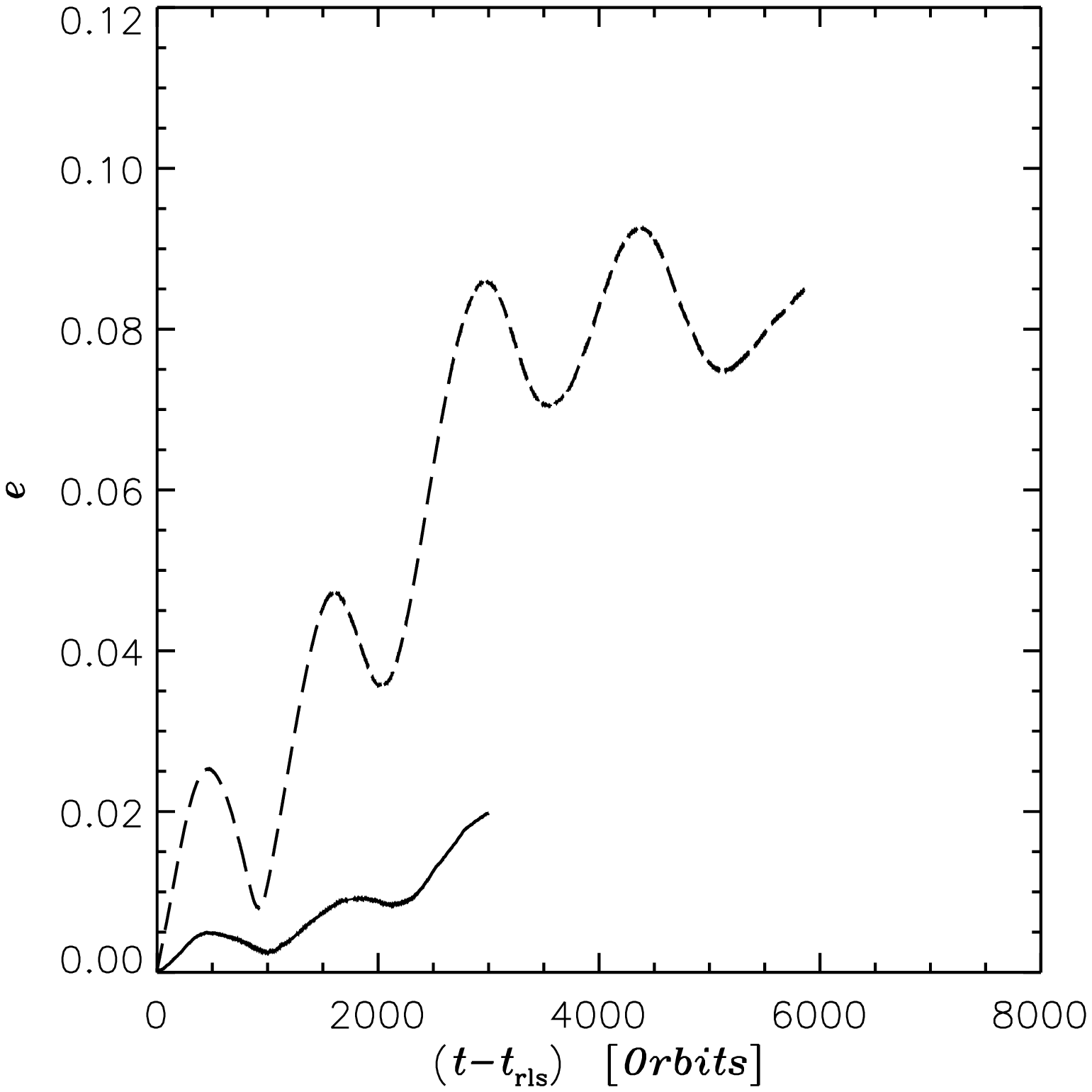}%
\includegraphics{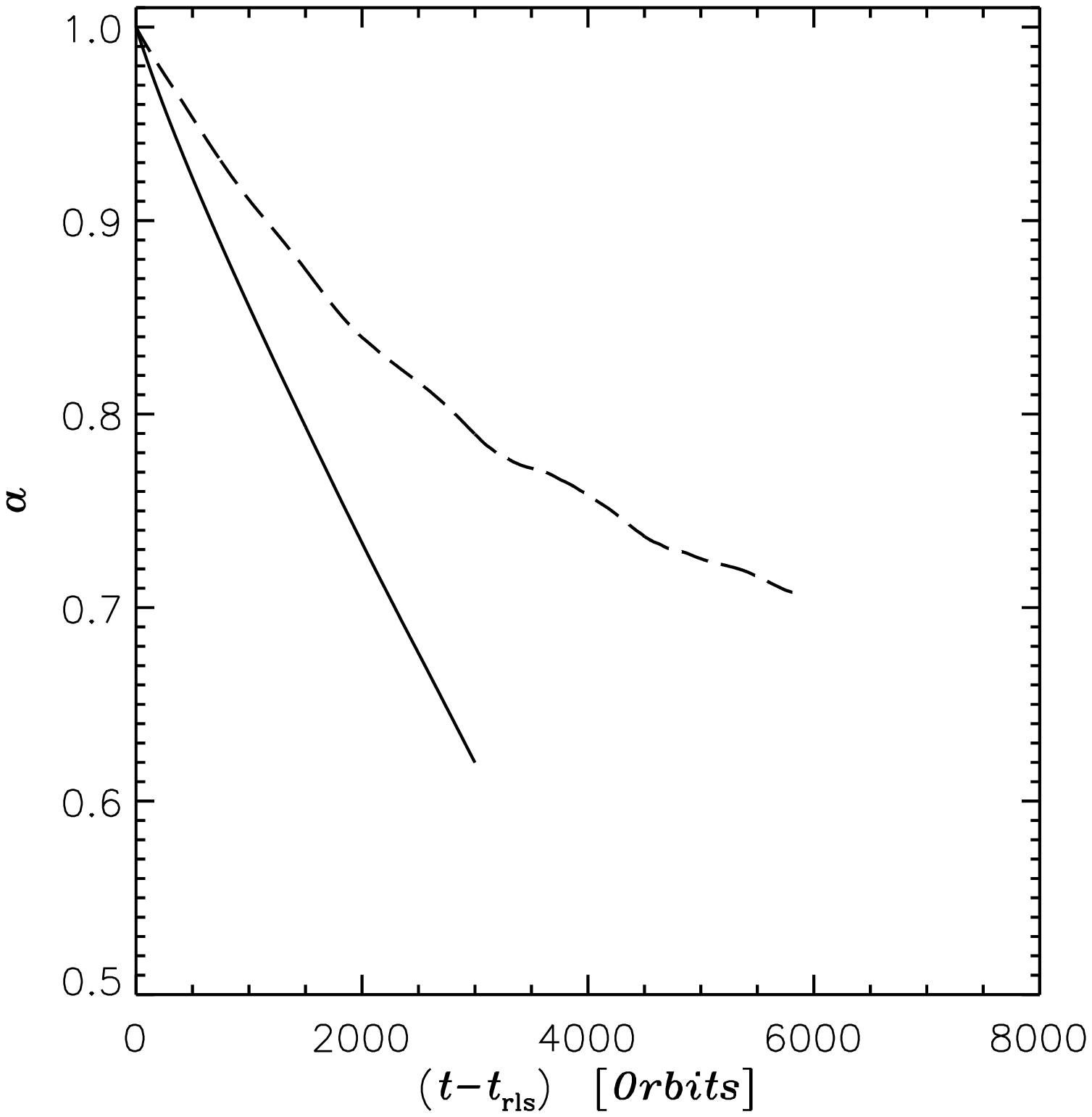}}
\caption{\small%
         Evolution of planet eccentricity (left) and semi-major axis 
         (right) for $3\,\MJup$ (top) and $1\,\MJup$ (bottom) planets 
         in disks with different values of viscosity parameter $\alpha_{0}$  
         ($\alpha$ at $r=a_{0}$),  $\alpha_{0}=1.2\times10^{-3}$ (long-dashed 
         line), $\alpha_{0}=4\times10^{-3}$ (solid line), and
         $\alpha_{0}=1.2\times10^{-2}$ (short-dashed line).
         The release time is $t_{\mathrm{rls}}=1000$ orbits
         in the top panels and $1100$ orbits bottom panels.
        }
\label{fig:eat_mp3vsnu}
\end{figure*}
As we discussed in \refSect{sec:viscosity_influence}, the disk eccentricity 
decreases with viscosity. For a coupled disk-planet system, we similarly 
expect that the planet eccentricity would  decrease with $\alpha$, since 
the eccentric corotation resonances become stronger. The orbital 
eccentricity evolution of  $3\,\MJup$ planets in  disks with different 
$\alpha$ values is shown in the top-left panel of \refFgt{fig:eat_mp3vsnu}.
The model with standard viscosity (solid line) is the same as that in 
\refFgt{fig:et_mp23}. Over a period of $2200$ orbits, the orbital 
eccentricity of the model with $\alpha_{0}=1.2\times10^{-2}$ 
(short-dashed line) remains small and never exceeds $e\approx0.01$. 
On the other hand, the models with lowest viscosities, 
$\alpha_{0}=1.2\times10^{-3}$ (long-dashed line) and $4\times10^{-3}$ 
exhibit a generally growing eccentricity.
The trend towards faster growth for smaller viscosities is confirmed by 
the model with $1\,\MJup$ and $\alpha_{0}=1.2\times10^{-3}$, as indicated 
by the long-dashed line in the bottom-left panel of \refFgt{fig:eat_mp3vsnu}.

The orbital evolution of the semi-major axis of $3\,\MJup$ planets for
the three disk viscosities is displayed in the top-right panel of 
\refFgt{fig:eat_mp3vsnu}. The radial inward migration is faster for 
larger $\alpha$, as expected in a Type~II-like regime. 
The plot supports the contention that planet eccentricity slows migration.
Around $4000$ orbits after the release time, the two calculations with 
smallest viscosities ($\alpha_{0}=4\times10^{-3}$ and $1.2\times10^{-3}$) 
produce migration rates respectively equal to 
$\dot{a}\approx-8\times10^{-6}\,a_{0}$ ($e\approx 0.11$) and 
$\dot{a}\approx-3\times10^{-6}\,a_{0}$ ($e\approx 0.14$) per orbit. 
The first rate is a factor $8$ smaller, while the second a factor $13$
smaller, than the initial migration speed (i.e., when $e\approx 0$).
For these cases, the average eccentricity growth rate at a time of about 
$2500$ orbits is $\dot{e}\approx 2\times10^{-5}\,\mathrm{orbit}^{-1}$.
The migration of a $1\,\MJup$ planet (\refFgp{fig:eat_mp3vsnu}, bottom-right 
panel) also shows that, as the orbital eccentricity approaches $\sim 0.08$, 
$|\dot{a}|$ starts to reduce. When $\alpha_{0}=1.2\times10^{-3}$, the average 
migration rate, over the last $1000$ orbits, is about a factor $5$ smaller 
than it is during the first $1000$ orbits of evolution after release.

\section{Pulsed Accretion}
\label{sec:planet_accretion}
\begin{figure*}
\centering%
\resizebox{0.95\linewidth}{!}{%
\includegraphics{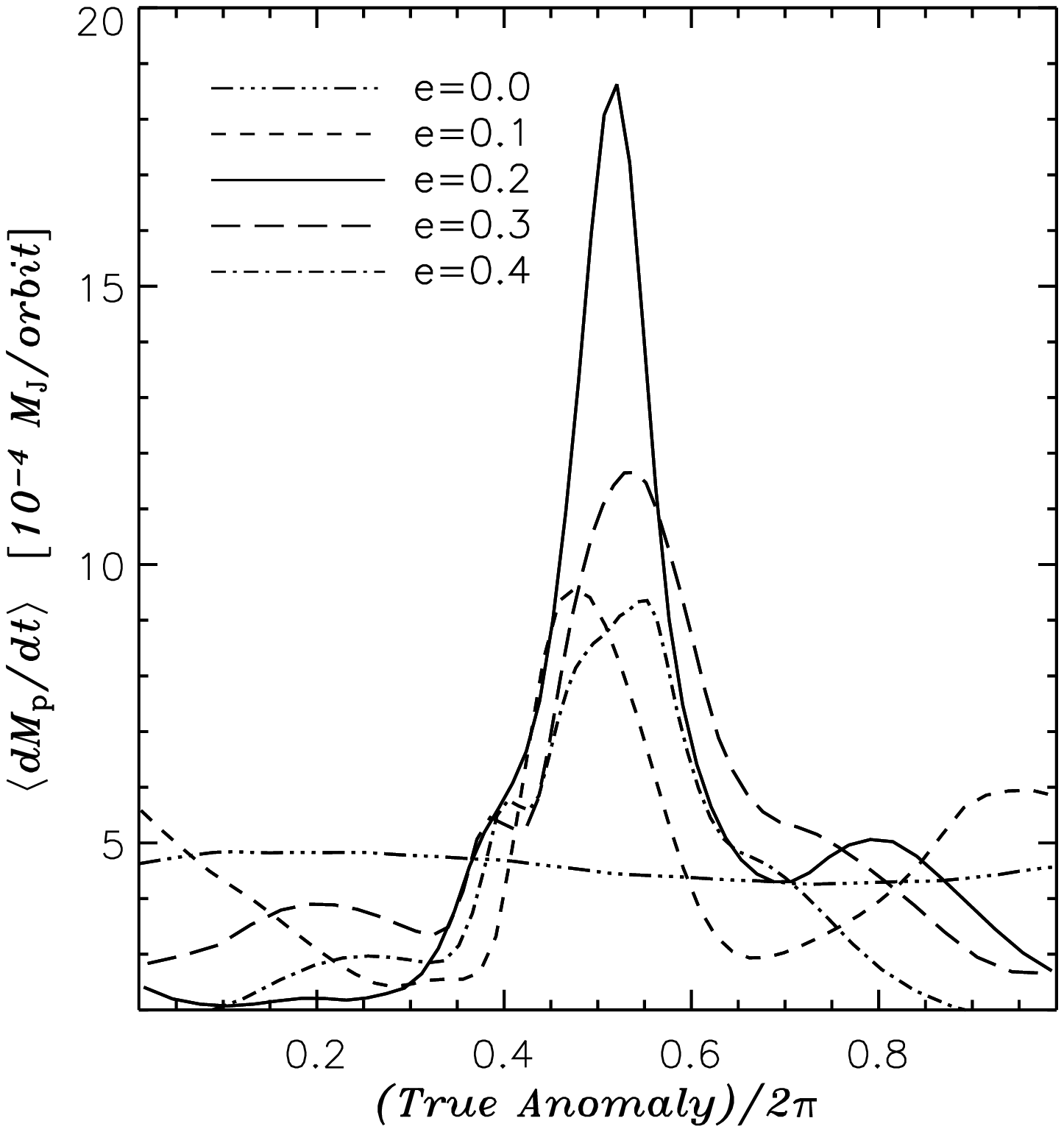}%
\includegraphics{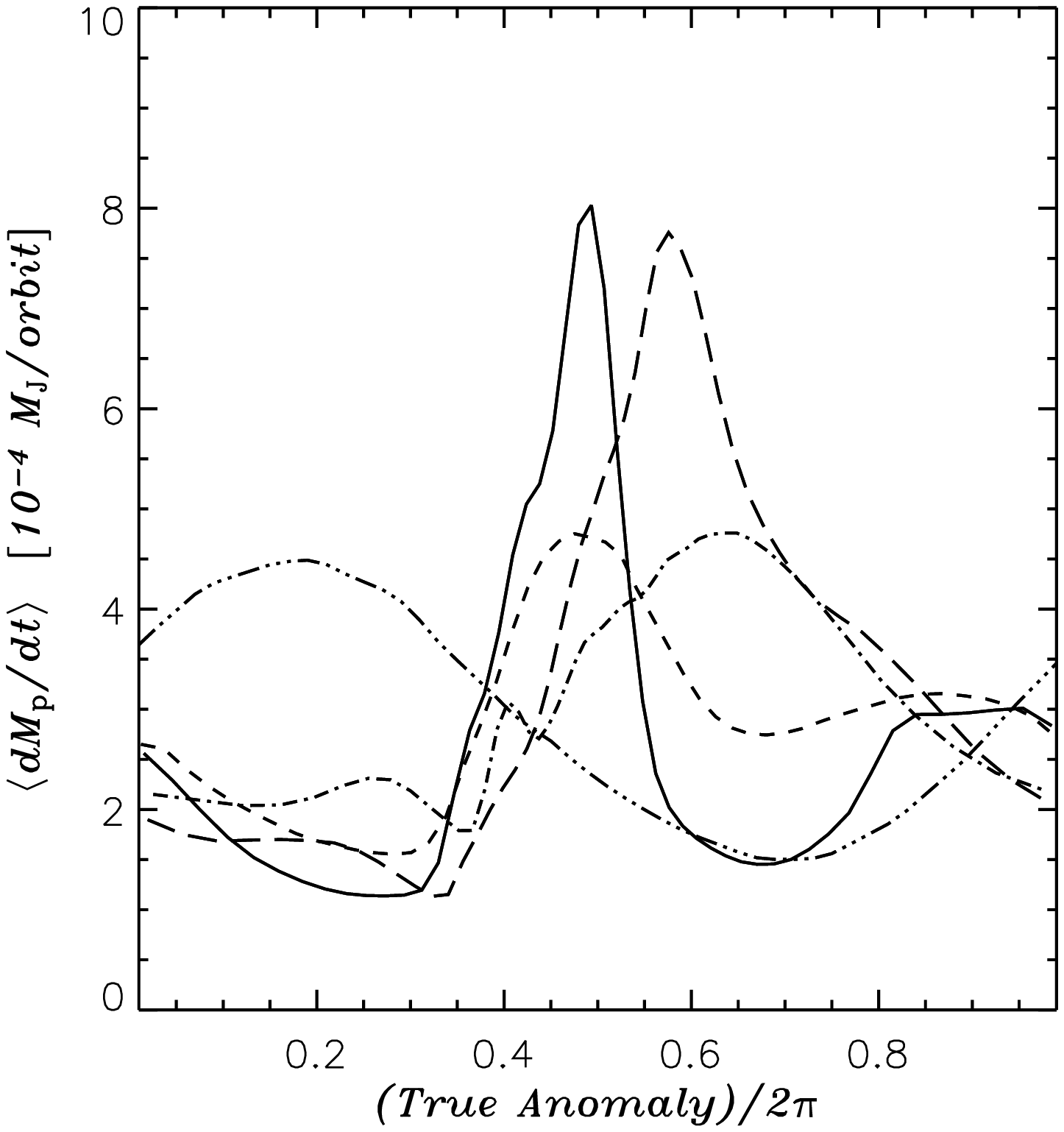}}
\caption{\small%
         Mass accretion rate in inner parts of the Roche lobe of a 
         $1\,\MJup$ (left) and a $3\,\MJup$ (right) planet as a function 
         of its true anomaly and orbital eccentricity.
         When the true anomaly equals $\pi$, the planet is located at the 
         apocenter of the orbit.
         Orbital eccentricities are listed in the legend of the left panel.
         The quantity $\langle\dMp\rangle$ is obtained by sampling $\dMp$
         every $0.02$ orbits and averaging the outcome from $500$ to $1000$ 
         orbits. 
        }
\label{fig:avmpdot}
\end{figure*}
For eccentric orbit binary star systems, the accretion from a
circumbinary disk onto the stars pulsates over the orbital period of 
the binary \citep{pawel1996,guenther2002}.
This effect is related to the pulsating character of the equipotential 
surfaces of the elliptical restricted three-body problem \citep{todoran1993}.

We measured the mass accretion rate  (which we denote as $\dMp$) into 
the inner portion of the planet's Roche lobe (within $\racc=0.3\,\Rhill$), 
following the prescription described in \refSect{sec:accretion_procedure}, 
as a function of the planet's orbit phase. We refer to this rate as the 
planet accretion rate although the flow is not resolved on the scale of 
the planet's radius and thus the rate at which the planet would accrete 
mass may be modulated somewhat differently from $\dMp$.
Quantity $\dMp$ was determined by folding the mass accretion rate over 
planet orbital phase and averaging over $500$ orbital periods (from $500$ 
to $1000$). \refFgt{fig:avmpdot} shows the resulting averaged accretion 
rate, $\langle\dMp\rangle$, for $1\,\MJup$ and $3\,\MJup$ planets, versus 
the true anomaly (i.e., the azimuthal position relative to pericenter) of 
the planet and as a function of the orbital eccentricity. The simulations 
show pulsed accretion in cases of eccentric orbit planets. The amplitude 
of the variability, and to a lesser extent the phase,  of $\dMp$ depends 
on the orbital eccentricity. 

\subsection{Modulation}
\label{sec:accretion_modulation}
The accretion onto a $1\,\MJup$ planet with $e=0.1$ has two asymmetric 
peaks, the taller of which is around the apocenter position (true anomaly 
equal to $\pi$). 
The secondary peak is  about $70$\% of the primary peak and is located close 
to the pericenter position. For $e \le 0.2$, the modulation of $\dMp$ 
increases with $e$. For larger orbital eccentricities, modulation decreases.
This effect may be a consequence of the gap becoming broader and shallower 
with increasing $e$ (see \refFgp{fig:avsigma}).

A similar phase variability is found for the mass accretion onto a $3\,\MJup$ 
planet (\refFgt{fig:avmpdot}, right panel). Even when the planet's orbit 
is circular, $\langle\dMp\rangle$ smoothly varies between 
$1.5\times10^{-4}\,\MJup$ and $4.5\times10^{-4}\,\MJup$ per orbit. 
The phasing in this case is arbitrary, since the planet's orbit is circular.
This behavior is related to the eccentricity of the disk. The mass accretion 
is markedly peaked around the apocenter position when $e>0.1$. The mass 
accretion modulation is again greatest for $e=0.2$.
When the planet eccentricity is between $e=0.3$ and $0.4$, the highest 
accretion rate occurs roughly $0.1$ orbits after apocenter. This delay 
may be related to the time required by material to be captured once it 
has been perturbed near the apocenter. Due to such a delay, accretion 
on binaries occurs near pericenter \citep{pawel1996}.

\begin{figure*}[t!]
\centering%
\resizebox{0.9\linewidth}{!}{%
\includegraphics{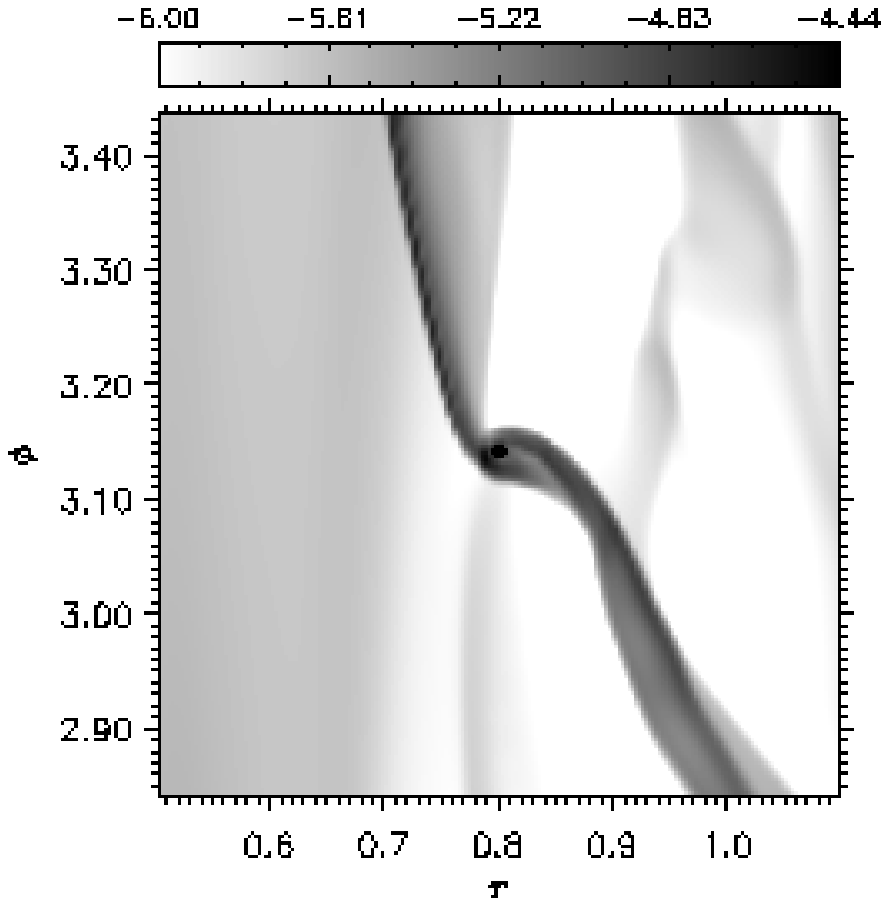}%
\includegraphics{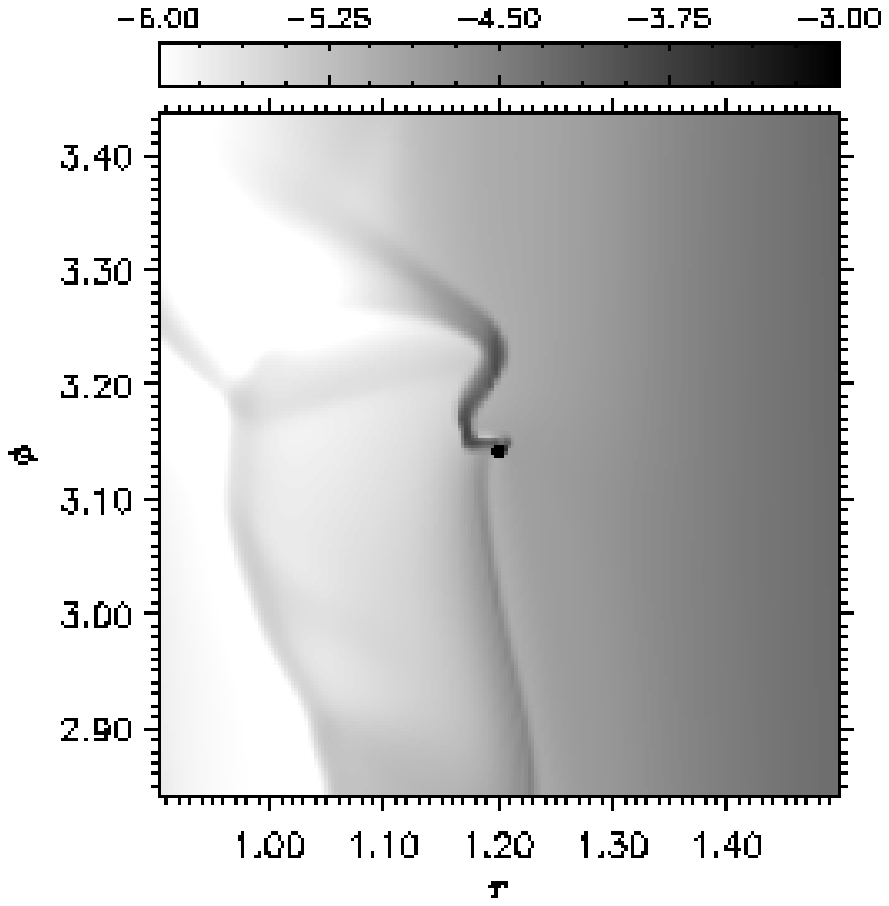}}
\resizebox{0.9\linewidth}{!}{%
\includegraphics{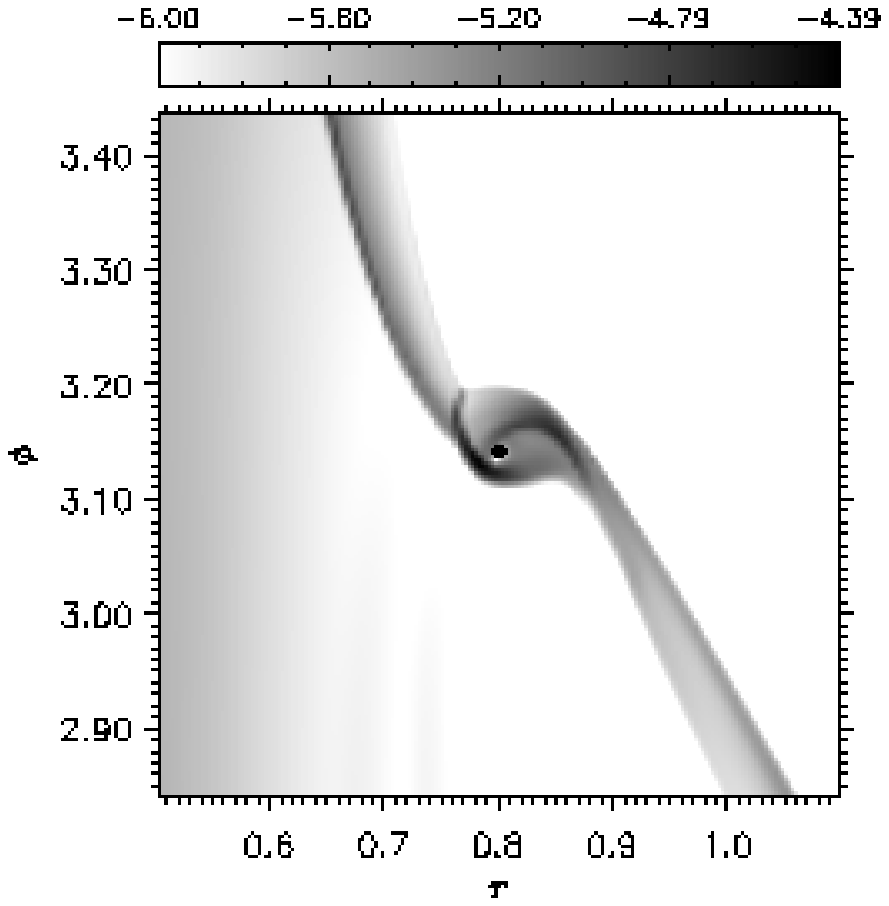}%
\includegraphics{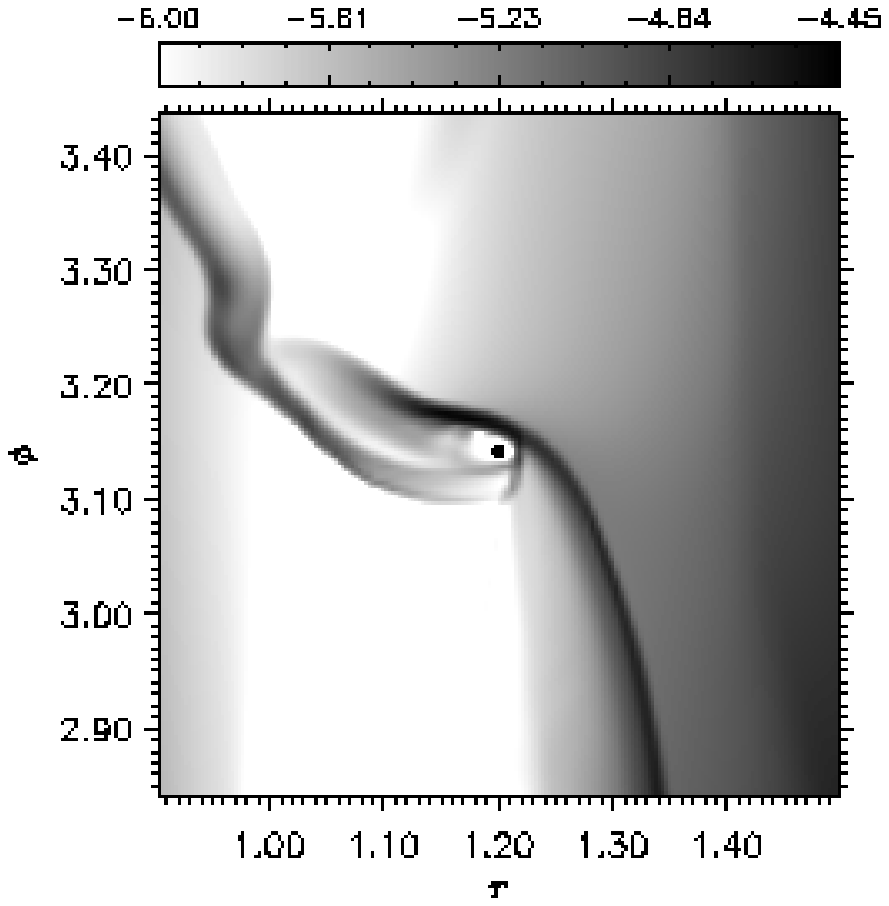}}
\caption{\small%
         Density structure around a $1\,\MJup$ (top) and a $3\,\MJup$ 
         (bottom) planet at pericenter  (left) and apocenter (right).
         The vertical axis is the azimuth about the star and the 
         horizontal axis is the distance from the star in units of 
         $a_{0}$. The orbital eccentricity is $e=0.2$ and $t\simeq 1000$ 
         orbits. The solid circle indicates the instantaneous location 
         of the planet. A surface density of $10^{-5}$ corresponds to 
         $3.29\,\sdunits$.
        }
\label{fig:density_zoom}
\end{figure*}
The density distribution in the vicinity of an eccentric orbit planet 
varies strongly with its orbital phase. This variation is illustrated 
in the panels of \refFgt{fig:density_zoom}, which depict the situation 
at the pericenter (left) and apocenter (right), for a $1\,\MJup$ (top) 
and a $3\,\MJup$ planet (bottom) on an eccentric orbit with $e=0.2$.
The spiral waves have a regular pattern at pericenter, when the planet 
is orbiting in the low-density gap (or cavity). As the planet approaches 
the apocenter, the outer spiral wake penetrates higher density regions, 
which causes fluid elements along the wake to lose angular momentum and 
flow through the gap. There are streams of material that extend inwards 
(at $r<a$ and $\phi>\phip$) which appear in the right panels of 
\refFgt{fig:density_zoom}.

\subsection{Mass Growth Timescale}
\label{sec:growth_time-scale}
The mass accretion rate onto a planet with a fixed circular orbit 
decreases with increasing planet mass when 
$\Mp\gtrsim 1\,\MJup$ \citep{lubow1999}. The average accretion rate in 
the simulations, at $t\simeq150$ orbits, of a $2\,\MJup$ planet is $0.63$ 
times that of a $1\,\MJup$ planet. The ratio decreases to $0.44$ for a 
$3\,\MJup$ planet on a circular orbit and at $t=150$ orbits. These ratios 
agree within $10\%$ with the values given by \citet{lubow1999}, who used
an independent code.

As the disk eccentricity grows, the mass accreted during an orbit increases. 
At later times $t\gtrsim 500$ orbits, when $S_{(1,0)} \gtrsim 0.2$, the 
accretion rate onto a $2\,\MJup$ planet  is $0.71$ times the rate onto a 
$1\,\MJup$ planet and for a $3\,\MJup$ planet  is $0.65$ times the rate 
onto a $1\,\MJup$ planet. Notice that, for a $3\,\MJup$  planet, this 
implies a $48$\% increase over the accretion rate at early stages, when 
the disk is circular. These results suggest that the eccentricity driven 
in the disk by a massive planet can augment the mass accretion rate onto 
the planet and hence shorten its growth timescale. 

Mass accretion over an orbit period can also be enhanced by the planet's 
orbital eccentricity. For a $1\,\MJup$ planet on a fixed orbit, the mass 
growth timescale (defined here as the ratio of $\Mp$ to the average 
accretion rate between $500$ and $1000$ orbits) decreased by about $35$\%
when $e$ is increased from $0$ to $0.2$. The reduction of mass growth 
timescale from $e=0$ to $e=0.4$ is only $17$\%, perhaps as a result of 
the wider gap and its smoother outer edge  at larger orbital eccentricities 
(see \refFgp{fig:avsigma}). For a $3\,\MJup$ planet on a fixed orbit with 
$e=0.3$, the mass growth timescale is reduced relative to the $e=0$ case by 
$27$\%. While for $e=0.4$ there was  a $13$\% reduction. For cases with 
$e < 0.3$, the growth rate was not substantially different than the $e=0$ 
case.

The mass growth timescale of a $1\,\MJup$ and a $3\,\MJup$ planet on 
circular orbit was also estimated for different values of the disk viscosity.
For a $3\,\MJup$ planet an increase in $\alpha$ by a factor of $3$ over the 
standard value (see \refSecp{sec:physical_par}) reduced the growth timescale 
by about $60$\%.
For both planetary masses,
a decrease in $\alpha$ by a factor of $3.3$ below the standard value, 
lengthened the growth timescale by a factor $2$.

Planetary accretion rates were determined by means of the grid system
GS2. Simulations executed with grid systems GS1 and GS3 resulted in very
similar outcomes (within $6$\%) for both the modulation of $\dMp$ and its 
average value. The results are not sensitive to the smoothing length, 
$\varepsilon$, although this parameter can affect the small-scale structure 
of the flow around the planet. We performed a calculation for an 
$\varepsilon$ value that was reduced  by a $25$\%, with $\Mp=3\,\MJup$ and 
$e=0.3$, and obtained essentially the same accretion rates (within $1$\%).

\subsection{Accretion towards the Star}
\label{sec:star_accretion}
\begin{figure}[t!]
\centering%
\resizebox{1.0\linewidth}{!}{%
\includegraphics{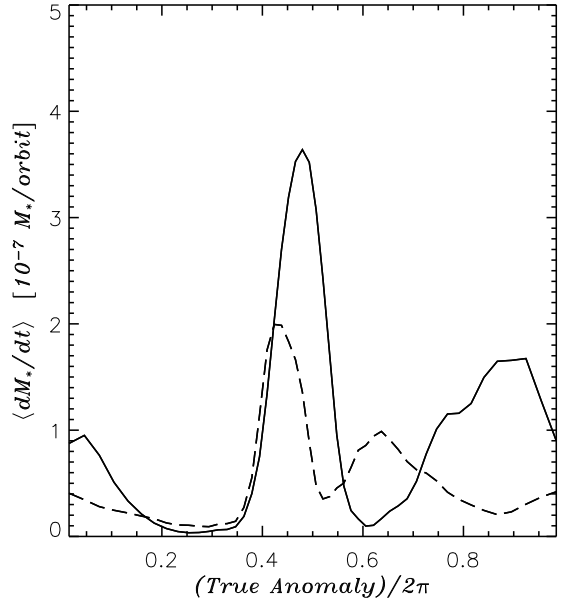}}
\caption{\small%
         Mass accretion rate towards the star at the inner boundary 
         $r=0.3\,a_0$ as function the true anomaly of the planet. 
         The mass accretion rate is averaged over several planet 
         orbital periods at $t\approx1000$ orbits. The dashed and solid 
         lines refer to models with $\Mp=1\,\MJup$ and $3\,\MJup$, 
         respectively. The planet's orbital eccentricity is $e=0.2$. 
         When the true anomaly is $\pi$, the planet is located at 
         the apocenter.
        }
\label{fig:avmsdot}
\end{figure}
The region interior to a planet's orbit likely contains an inner disk 
which cannot be resolved in the current two-dimensional simulations 
(see discussion in \refSecp{sec:radial_migration}). We estimate the 
modulation of mass onto this disk as a function of orbital phase by 
considering the mass flow rate across the inner boundary. As material
accretes through the inner disk, the modulation would be expected to weaken. 
It is unclear whether the modulation would be reflected as a variable 
accretion at the surface of the star. Perhaps it could be manifested 
as variability in emission from the region where the inflow meets the 
outer edge of the inner disk.

\refFgt{fig:avmsdot} plots the accretion rate at the inner boundary 
$\langle\dMs\rangle$ versus the true anomaly of the planet for cases 
with $\Mp=1\,\MJup$ and $3\,\MJup$. The case involving the more massive 
planet produces an accretion rate through the inner boundary that is 
largest when the planet is close to the apocenter. The maximum of 
$\langle\dMs\rangle$ occurs before the apocenter passage for the case 
involving the $1\,\MJup$ planet. The average mass accreted by the star 
during one orbital period of the planet is $5.8\times10^{-8}\,\MStar$ 
and $9.9\times10^{-8}\,\MStar$ for the $1\,\MJup$ and the $3\,\MJup$ 
cases, respectively. The same models with no orbital eccentricity show 
relatively constant $\langle\dMs\rangle$ values of 
$1\times10^{-7}\,\MStar$ ($\Mp=1\,\MJup$) and 
$6.3\times10^{-8}$ ($\Mp=3\,\MJup$) per orbit.
As a comparison, the mass accretion through a steady-state $\alpha$-disk
is $3\pi\Sigma\,\nu$ \citep{lynden-bell1974,pringle1981} which, using the
initial unperturbed surface density and standard viscosity 
($\alpha_{0}=4\times10^{-3}$), yields $2.5\times10^{-7}\,\MStar$
per orbit or $2\times10^{-8}\,\Msyr$.

The ratio of the accretion on the star to the accretion in the disk,
outside the gap, can be expressed as 
$\langle\dMs\rangle_{\mathrm{o}}/%
(\langle\dMs\rangle_{\mathrm{o}}+\langle\dMp\rangle_{\mathrm{o}})$,
where the subscript ``$\mathrm{o}$'' denotes is the integral  of 
the respective accretion rate over a planetary orbit.
For both the $1\,\MJup$ and the $3\,\MJup$ planets on circular orbit, this
ratio is $0.19$. When $e=0.2$, the ratio is $0.09$ and $0.26$
for the $1\,\MJup$ and the $3\,\MJup$ planet, respectively. 
The reduced mass transfer across the planet's orbit, when $\Mp=1\,\MJup$ 
and $e=0.2$, can be attributed to the increased accretion rate onto the 
planet, as reported above. In the  $3\,\MJup$ case, 
$\langle\dMp\rangle_{\rm o}$ does not vary significantly as $e$ varies 
from $0$ to $0.2$. Instead, the mass flux across the gap is likely enhanced 
by the radial excursion of the planet (see \refSecp{sec:planet_accretion}).

\section{Summary and Discussion} 
\label{sec:conclusions} 
We simulated the orbital evolution of circular and eccentric orbit giant  
planets embedded in circumstellar disks.  The disks were  analyzed using  
a two-dimensional hydrodynamics code that utilizes nested grids to achieve  
high resolution in a large region ($2\,a\times 2\pi/3$) around the  
planet. The disks were modeled as an $\alpha$-disk and a few values of  
$\alpha$ were considered. We investigated planet masses of $1\,\MJup$,  
$2\,\MJup$, and $3\,\MJup$ and initial orbital eccentricities that ranged  
from $0$ to $0.4$. 
 
Disk gaps become broader and shallower as the planet eccentricity increases 
(see \refFgp{fig:globalsigma} and \ref{fig:avsigma}). The density near the  
orbit of the planet is very small compared with the density in the disk for  
all eccentricities considered. A planet on a fixed circular orbit can cause  
an initially circular disk to become eccentric (see \refFgp{fig:rcmvstime}).  
The disk eccentricity is suppressed at lower planet masses  
($\Mp\lesssim 1\,\MJup$) and higher disk viscosities ($\alpha\gtrsim 0.01$),  
as also found by \citet{kley2006} and by \citet{papa2001} at higher planet  
masses. We attribute the eccentricity growth to a tidal instability  
associated with a series of eccentric outer Lindblad resonances in the inner  
parts of the outer disk (\refFgp{fig:lam}). The same type of instability,  
involving an inner disk, is thought to be responsible for the superhump  
phenomena in binary star systems \citep{lubow1991a,osaki2003}.  
 
The simulations indicate that planet eccentricity can grow, as a consequence  
of disk-planet interactions (\refFgp{fig:et_mp23} and \ref{fig:et_mp1vse}).  
The growth is stronger in the $2\,\MJup$ and $3 \,\MJup$ cases than for  
$1\,\MJup$, and for lower disk viscosity ($\alpha\lesssim 4\times10^{-3}$).  
Planet eccentricities of $\sim 0.1$ were found in the simulations over the  
course of a few thousand orbits for $2 \,\MJup$ and $3 \,\MJup$  planets.  
A similar eccentricity growth is obtained for a $1 \,\MJup$ planet in a 
disk with viscosity $\alpha\approx 10^{-3}$. 
The planet and disk 
both acquire eccentricity as they interact, which may lead to complicated  
time-dependent behavior of their eccentricities. The planet eccentricity 
growth is likely aided by the disk eccentricity growth. The results suggest  
that the eccentric growth found for $\sim 10\,\MJup$ planets by  
\citet{papa2001} also occurs for lower planet masses. 
The higher resolution achieved by our calculations may be playing a role  
in obtaining this growth. 
 
For circular orbit planets, 
migration occurs on roughly the local viscous timescale, as expected for 
Type~II migration. However, it is slowed for eccentric orbit planets. 
This result appears for several configurations with either dynamically 
determined (\refFgp{fig:at_mp123} and \ref{fig:eat_mp3vsnu}) or imposed 
planet eccentricities  (\refFgp{fig:at_mpvse}). For a $2\,\MJup$ case, 
even migration reversal (outward migration) is found for a dynamically 
determined eccentricity (\refFgp{fig:at_mp123}).
Migration slowing or reversal would have important consequences for the 
planet formation process.
The cause is not yet clear. It may involve torques from outer disk 
\citep{papa2002} or instead from the coorbital region. Some preliminary 
evidence suggests the latter.

Mass accretion both within a planet's Roche lobe and through a gap can be 
strongly modulated with orbital phase for eccentric orbit planets or 
eccentric disks (\refFgp{fig:avmpdot}, \ref{fig:density_zoom}, and 
\ref{fig:avmsdot}). The modulation was largest for planet eccentricity 
$e \simeq 0.2$. This pulsating accretion is similar to what is found for
eccentric orbit binary stars embedded in a circumbinary disk 
\citep{pawel1996}, although the phasing is different. Both disk and planet 
eccentricity also lead to enhanced accretion onto the planet. 
This enhancement likely helps planets achieve higher masses.

The simulations lend support to the idea that disk-planet interactions 
cause planet eccentricity growth, along the lines of \citet{goldreich2003}. 
The simulations suggest that planet eccentricities are easier to achieve 
for higher mass planets ($\Mp\gtrsim 2\,\MJup$).
Our results are subject to the usual
limitations in approximate initial conditions, simulation time, radial 
range for coverage of the disk (likely resulting in the lack of an inner 
disk), the $\alpha$-disk model, and the use of various numerical devices. 
We also neglected disk self-gravity, which may affect migration especially 
for higher mass disks \citep{anelson2003a}.

However, it is not clear that typical extra-solar planet eccentricities 
of $0.2$--$0.3$ can be achieved through disk-planet interactions. 
The eccentricity growth at later times shows indications of slowing and 
possibly stalling for $e \la 0.15$ (see Fig~\ref{fig:eat_mp3vsnu}.)
Perhaps higher eccentricities can be achieved for disks with
different properties (e.g., lower viscosity and smaller disk's 
aspect ratio).
Eccentricity may be limited by damping due to high order eccentric inner 
Lindblad resonances that lie outside a planet's orbit. Simulations of 
eccentric orbit binary star systems suggest that little eccentricity 
growth occurs for $e \ga 0.5$ \citep{lubow1992b}. 
Although the simulated planets do not achieve orbital eccentricities in 
excess of $0.15$ over the duration of the simulated evolution (for 
configurations that start from circular orbits), the simulation times 
correspond to less than $10^{5}$ years. Migration slowing and reversal
may permit the planets to achieve higher eccentricities on longer 
timescales while avoiding orbit decay into the disk center/host star.

We have not yet investigated eccentricity evolution of sub-Jupiter mass 
planets. They may also provide important constraints. Other simulations
suggest that disks with standard viscosity have only mild gaps for smaller 
planet masses of $\Mp \la 0.1\,\MJup$ \citep[e.g.,][]{gennaro2003b,bate2003}.
Under those conditions, disk-planet interactions likely lead to eccentricity 
damping, due to the dominance of the coorbital Lindblad resonance
\citep{ward1986,pawel1993}. 
The observational determination of eccentricities for small 
mass planets would help constrain these models. The planet around HD~49674 
is close to this regime. It has a minimum mass of $0.11\,\MJup$ and a 
best-fit eccentricity of $0.29$ (P.~Butler, private communication). Since 
it is close to the central star (the period is $4.9$ days), it is possible 
that the eccentricity evolution could be more complicated, especially if 
it became trapped in a central disk hole. Examples of isolated planets like 
this, but at longer periods would provide useful constraints.



\acknowledgments

We thank Gordon Ogilvie and Jim Pringle for useful discussions.
The computations reported in this paper were performed using
the UK Astrophysical Fluids Facility (UKAFF).
GD was supported by the Leverhulme Trust through a UKAFF Fellowship, 
by the NASA Postdoctoral Program, and in part by NASA's Outer Planets 
Research Program through grant 811073.02.01.01.20.
SL acknowledges support from NASA Origins of Solar Systems grant
NNG04GG50G.
MRB is grateful for the support of a Philip Leverhulme Prize.




\end{document}